\documentclass[12pt]{article}
\usepackage{amsthm,amsmath,latexsym,amssymb,amsfonts,amscd}
\usepackage{graphicx,lscape,fancyhdr,array,stmaryrd,euscript,wrapfig}
\usepackage{ifthen,empheq}
\usepackage{sidecap}
\usepackage{rotating,framed}
\usepackage{ifpdf}
\usepackage{verbatim}
\usepackage{color}

\numberwithin{equation}{section}

\usepackage{tikz}
\usepackage{hyperref}
\usepackage[
	sorting=none,
	maxnames=99,
	firstinits=true,
	doi=false,
	citestyle=numeric-comp,
	backend=bibtex
	]{biblatex}

\renewbibmacro{in:}{}
\def\sideremark#1{\ifvmode\leavevmode\fi\vadjust{\vbox to0pt{\vss
 \hbox to 0pt{\hskip\hsize\hskip1em
 \vbox{\hsize1.5cm\tiny\raggedright\pretolerance10000
 \noindent #1\hfill}\hss}\vbox to8pt{\vfil}\vss}}}%

\newcolumntype{x}[1]{%
>{\centering\hspace{0pt}}m{#1}}%
\newcolumntype{w}[1]{%
>{\raggedright\hspace{0pt}}m{#1}}%
\newcolumntype{z}[1]{%
>{\raggedleft\hspace{0pt}}m{#1}}%

\newcommand{\pl}{\partial}
\newcommand{\be}{\begin{equation}}
\newcommand{\ee}{\end{equation}}

\newcommand{\mm}{{\ensuremath{\underline{m}}}}
\newcommand{\nn}{{\ensuremath{\underline{n}}}}

\newcommand{\gad}{{\dot{\alpha}}}
\newcommand{\gbd}{{\dot{\beta}}}

\newcommand{\gnd}{{\dot{\nu}}}

\newcommand{\ga}{\alpha}
\newcommand{\gb}{\beta}
\newcommand{\gc}{\gamma}

\newcommand{\bry}{{{\bar{y}}}}

\newcommand{\brxi}{{{\bar{\xi}}}}

\newcommand{\breta}{{{\bar{\eta}}}}
\newcommand{\brzeta}{{{\bar{\zeta}}}}

\newcommand{\brn}{{{\bar{n}}}}
\newcommand{\brm}{{{\bar{m}}}}

\newcommand{\fud}[2]{{}^{#1}{}_{#2}\,}
\newcommand{\fdu}[2]{{}_{#1}{}^{#2}\,}

\newcommand{\pC}{{C}} 
\newcommand{\adD}{{D}} 
\newcommand{\tadD}{{\widetilde{D}}} 

\newcommand{\formJ}{{\boldsymbol{\mathsf{J}}}}

\newcommand{\formK}{{\boldsymbol{\mathsf{K}}}}
\newcommand{\formU}{{\boldsymbol{\mathsf{U}}}}

\newcommand{\besubeqs}{\begin{subequations}}
\newcommand{\esubeqs}{\end{subequations}}

\begin{document}
\pagestyle{empty}
\hfill
\vskip 0.1\textheight
\begin{center}

{\Large\bfseries On Locality, Holography and Unfolding} \\

\vskip 0.1\textheight

Evgeny \textsc{Skvortsov}$\,{}^{\sharp,\flat}$ and Massimo \textsc{Taronna}${}^{\sharp}$

\vspace{2cm}

{\em ${}^{\sharp}$  Albert Einstein Institute, Golm, Germany, D-14476, Am M\"{u}hlenberg 1}\\
\vspace*{5pt}
{\em ${}^{\flat}$ Lebedev Institute of Physics, Leninsky ave. 53, 119991 Moscow, Russia}

\end{center}

\vskip 0.05\textheight

\begin{center}
	{\bf Abstract }

\end{center}
\begin{quotation}
\noindent  We study the functional class and locality problems in the context of higher-spin theories and Vasiliev's equations. A locality criterion that is sufficient to make higher-spin theories well-defined as field theories on Anti-de-Sitter space is proposed. This criterion identifies admissible pseudo-local field redefinitions which preserve AdS/CFT correlation functions as we check in the 3d example. Implications of this analysis for known higher-spin theories are discussed. We also check that the cubic coupling coefficients previously fixed in 3d at the action level give the correct CFT correlation functions upon computing the corresponding Witten diagrams.
\end{quotation}

\newpage
\pagestyle{plain}

\tableofcontents

\section{Introduction}

Higher-spin (HS) theories are receiving a great amount of attention in the recent years (see \emph{e.g.} \cite{Bekaert:2005vh,Bekaert:2010hw,Taronna:2012gb,Giombi:2012ms,Gaberdiel:2012uj,Sagnotti:2013bha,Didenko:2014dwa} for some reviews). Both in 3d and 4d, key conjectures have been put forward linking Vasiliev's equations \cite{Vasiliev:1990en,Prokushkin:1998bq} to free or interacting vectorial CFTs \cite{Sezgin:2002rt,Klebanov:2002ja,Gaberdiel:2010pz}. The power of these dualities relies in their weak-weak character, to be distinguished from the standard manifestations of holographic dualities. In principle this feature allows to directly compare bulk and boundary physics, allowing to test quantum gravity and holography itself. These conjectures, mostly based on the analysis of symmetries of the boundary theories, have driven a lot of interest also in relation to string theory \cite{Bianchi:2003wx,Engquist:2005yt,Taronna:2010qq,Sagnotti:2010at,Gaberdiel:2014cha} with a resurgence of old and new relations among string theory and Vasiliev's equations (see \emph{e.g.} \cite{Giombi:2011kc,Chang:2012kt}). So far, most of the attention has been limited to the CFT side, where we currently dispose of many field theory tools that allow remarkable computations and checks within a standard and well understood setting. On the contrary, the bulk Vasiliev side has attracted less attention beyond the free theory regime, with the exception of a handful of works \cite{Sezgin:2002ru,Sezgin:2003pt,Giombi:2009wh,Giombi:2010vg,Chang:2011mz,Ammon:2011ua,Kessel:2015kna,Boulanger:2015ova}. A better understanding of the bulk side of the correspondence is however desirable in order to achieve a better grasp of Vasiliev's equations and also of the proposed dualities. 

Many questions are still to be addressed and in this paper we aim to consider the problem of defining an appropriate functional class within the unfolded formalism \cite{Vasiliev:1988xc,Vasiliev:1999ba} to describe what one would call a local bulk theory in this framework (see \emph{e.g.} \cite{Vasiliev:2015wma} for some recent attempts in this direction). We also underline the links between the bulk and boundary side of the problem towards a better grasp of locality from an AdS/CFT perspective.

Before moving to the actual analysis, it is important to recall that HS theories are based on an infinite-dimensional symmetry, HS symmetry \cite{Fradkin:1986ka}, that at the free level is given by a HS algebra acting on the jet space of fields. The HS algebra often admits a $\star$-product realisation \cite{Fradkin:1986ka,Vasiliev:2000rn,Vasiliev:2004cm} and combines fields of all spins $s=0,1,2,3,...$ into a single multiplet. Moreover, the HS algebra relates also different derivatives of the fields together in a single multiplet. In particular, a HS transformation can change the order in derivatives of a given expression and HS algebra covariant expressions contain for this reason infinitely many derivatives, i.e. are naturally pseudo-local (see \emph{e.g.} eq.~\eqref{standbackr}). This is the reason why understanding pseudo-local quantities is of utter importance for HS theories.

At present all examples of HS theories are given by Vasiliev's equations \cite{Vasiliev:1990en,Prokushkin:1998bq,Vasiliev:2003ev}. Vasiliev's equations are generating equations, i.e. they lead to the physical HS equations of motion provided they are partially solved (with respect to certain auxiliary directions) and, most importantly, the gauge is fixed appropriately.\footnote{Notice here that different gauge choices in this context correspond to different field frames related by pseudo-local field redefinitions after the equations have been solved with respect to the auxiliary directions.} Solving for the auxiliary directions in different gauges might well lead to inequivalent results if the corresponding gauge choices are inequivalent or, possibly, inconsistent. So far the most convenient gauge choice has been the Schwinger-Fock gauge, whose main virtue is to make the link between unfolded formalism and Fronsdal's equations more direct and, as well, to make manifest Lorentz symmetry at any perturbative order.

Remarkably, in the Schwinger-Fock gauge mentioned above, the unfolded HS equations that one can extract from the Vasiliev equations turn out to contain certain pseudo-local tails \cite{Kessel:2015kna,Boulanger:2015ova}. In particular, the second-order equations, as extracted from Vasiliev's theory, have pseudo-local sources on the r.h.s. of the Fronsdal equations:
\begin{equation}
(\square -m^2_\phi)\phi_{\mm_1...\mm_s}+...=\sum_{l,q} \alpha_{s,q}^l (\Lambda)^{-l}\,\underbrace{g^{\nn\nn}\cdots g^{\nn\nn}}_{l}\nabla_{\mm(s-q)\nn(l)}\Phi^* \nabla_{\mm(q)\nn(l)}\Phi+ \ldots\,. \label{standbackr}
\end{equation}
This is true surprisingly even at the quadratic- (cubic- from the point of view of the action) order, where it is well known that interactions are higher-derivative but local --- given three spins the number of independent vertices is finite while each of them have a number of derivatives not exceeding the sum over spins \cite{Metsaev:2005ar}. We refer to the latter couplings as \emph{primary canonical} couplings hereafter, to stress that they have the minimal number of derivatives and that improvement terms have been discarded.
The appearance of such expressions can be easily expected taking into account the choice of field frame selected by HS covariance in Vasiliev's equations.

Indeed, on the one hand, when looking for an action one begins with a quadratic Lagrangian that describes the propagation of free fields over a fixed background, of which the empty Anti-de-Sitter (AdS) space is the usual choice in the HS case.
On the other hand, when interactions are included, one can have access to more complicated HS backgrounds, of which a flat connection of the HS algebra is the simplest example, and empty AdS space is just a particular case. From this respect, unfolded HS equations are naturally an expansion over a flat connection of the HS algebra rather than just AdS space. Therefore, also the first nontrivial corrections to the free unfolded equations must contain some information about backgrounds that are more complicated than empty AdS space.
This feature is a direct analogue, in the context of HS theories, of the complexity of cubic couplings extracted from the Einstein-Hilbert action. Those contain indeed a particular combination of improvement terms required by general covariance. Analogously, the effect of HS covariance when extracting the couplings from Vasiliev's equations naturally manifests itself into pseudo-local tails with an infinite number of improvement terms. Such pseudo-local tails are therefore natural in view of the HS symmetry, which mixes derivatives of all orders and fields of all spins together.

In this paper we want to quantitatively address the question of locality in AdS space for a generic pseudo-local field theory at least to the lowest non-trivial order. The motivation is twofold in view of the above discussion: first of all, it is desirable to have a quantitative understanding of what it means to have a local theory in AdS, especially when one deals with pseudo-local expansions in derivatives; secondly, it is compelling to improve our understanding of the degree of (non-)locality of existing HS theories. This issue becomes relevant in view of the fact that one is also led to consider field-redefinitions that are pseudo-local. Allowing for generic pseudo-local redefinitions erases the difference between trivial and nontrivial interactions\footnote{Notice for instance the result of \cite{Prokushkin:1998bq} in which the interactions in 3d Vasiliev's equations were shown to be reconstructed via a pseudo-local redefinition expressed in terms of an integration flow.} \cite{Prokushkin:1998bq,Prokushkin:1999xq,Kessel:2015kna} and allows in principle also to solve the Noether procedure up to any order in perturbation theory without any obstruction\footnote{It can be useful to recall here the logic of \cite{Taronna:2011kt}. The logic there was first to study the generic non-local solution and then look for a locality criterion to identify a \emph{local} solution among the non-local ones. In flat space no local solution could be found. However, in our case there are various holographic indications that such local solution should indeed exist when expanding around AdS background.} \cite{Barnich:1993vg,Taronna:2011kt}. For these reasons the problem of constraining admissible pseudo-local redefinitions cries for a prescription of functional class to give them a well-defined physical interpretation.

Our proposal at this order can be summarised as follows. (i) first one decomposes a given coupling into an orthogonal basis made of primary canonical currents and improvements; (ii) one is allowed to redefine away each local improvement one by one; (iii) we strictly forbid pseudo-local redefinitions that remove primary canonical currents (\emph{e.g.} the usual stress tensor or the electromagnetic current). In this way the (pseudo-local) redefinition will never change the coefficient of the primary canonical component. Admissible redefinitions defined here can only touch improvement terms. Let us also stress that at this order, as we will prove in the following, there exists a basis of improvements and canonical currents such that each element of such basis is local.

While our locality proposals are bench-marked against concrete Vasiliev's HS theories, the considerations are quite general and apply to any theory in AdS. Finally, our main result can be summarised by the following conditions for the asymptotic behaviour of the coefficients $\alpha_{s,q}^l$ in eq.~\eqref{standbackr}:
\begin{subequations}
\begin{align}
\alpha_{s,q}^l&\prec \frac{1}{l^{2s}}\frac{1}{(2l)!}\,,&d&=3\,,\\
\alpha_{s,q}^l&\prec \frac{1}{l^{2s+1}}\left(\frac{1}{l!}\right)^2\,,&d&=4\,,
\end{align}
\end{subequations}
compatibly with a consistent resummation of the above higher-derivative tails.

The plan of the paper is as follows. In Section \ref{Sec: General Crit} we discuss the main logic of the paper. In Section \ref{Sec: Ambient} we discuss the locality criterion in the ambient space formalism and in generic dimensions. In Section \ref{Sec: 3d}, we analyse the same problem in the context of 3d unfolding. In Section \ref{Sec: 4d} we extend the analysis to 4d unfolding and in Section \ref{Sec: Witten Diagrams} we study the implications of the locality criterion from the AdS/CFT perspective. We discuss and summarise the results and their implications in Section \ref{Sec: Discussions}. Our conventions are listed in Appendix \ref{app:notation}.

\section{Redefinition and Functional Class: a General Criterion}\label{Sec: General Crit}

Before moving to detailed examples we would like to describe the general ideas and framework. In this section we illustrate the generic structure of pseudo-local couplings discussing our main proposal for a definition of functional class in a pseudo-local field theory. As we will see, the locality criterion proposed here will not be fulfilled by Vasiliev's equations in the Schwinger-Fock gauge, which appear to be a bit more non-local than expected. We will comment more on the consequences of this observation in the discussion section.

Without loss of generality, the essence of our discussion can be illustrated starting from the simplest possible cubic coupling describing the interaction of a scalar field with a spin-$s$ HS field:
\begin{equation}
\int  \, \phi_{\mm(s)} J^{\mm(s)}\,.
\end{equation}
This induces a source in the Fronsdal equations \cite{Fronsdal:1978rb} of the type:
\begin{equation}
\Box\phi_{\mm(s)}+\ldots=J_{\mm(s)}\,.
\end{equation}
Here $\phi_{\mm(s)}$ is a massless totally symmetric Fronsdal's field \cite{Fronsdal:1978rb} subject to the gauge invariance condition:
\begin{align}
\delta \phi_{\mm(s)}&= \nabla_{\mm} \xi_{\mm(s-1)}\,,& \xi_{\mm(s-3)\nn}{}^\nn&=0\,,
\end{align}
while $J^{\mm(s)}$ is a conserved rank-$s$ tensor that for example can be built out of two scalar fields as:
\begin{equation}
\label{standardcurrent}
J^{can}_{\mm(s)}\approx\Phi^* \overleftrightarrow{\nabla}_{\mm} ... \overleftrightarrow{\nabla}_{\mm}\Phi + \mathcal{O}(\Lambda g_{\mm\mm})\,.
\end{equation}
We call it hereafter \emph{primary} canonical current to distinguish it from its higher-derivative dressings below.
The $\mathcal{O}(\Lambda g_{\mm\mm})$ terms are proportional to the uncontracted metric $g_{\mm\mm}$ and are required in order to make the above currents conserved in AdS. Furthermore, by convention, the covariant derivatives are symmetrised and unless specified we set $\Lambda=1$.
While it is well known that the above current is unique modulo redefinitions, off-shell one can write down an infinite number of local improvements. The existence of infinitely many local improvements is tantamount to the possibility of considering higher-derivative dressings of the above couplings described by more general conserved currents that take the following form:
\begin{align}\label{nonl}
J^{pl}_{\mm(s)}&=\sum_{l,q} (\Lambda)^{-l}\alpha_{s,q}^l (g^{\nn\nn})^{l}\nabla_{\mm(s-q)\nn(l)}\Phi^* \nabla_{\mm(q)\nn(l)}\Phi+ \mathcal{O}(\Lambda g_{\mm\mm})\,.
\end{align}
We will refer to such higher-derivative currents ($l>0$) as \emph{successors} of the primary canonical one \eqref{standardcurrent} to stress that they have the same structure but dressed with a tail of derivatives. We call otherwise the sector of all currents proportional to the primary canonical one with possible higher derivative dressings just \emph{canonical}.

Assuming for a moment to consider such pseudo-local tails in Minkowski space, in terms of some other dimensionful parameter like $\alpha^\prime$ but keeping massless the HS fields, we would quickly discover that those terms boil down to something vanishing. The main reason is due to commutativity of derivatives. Indeed in flat space one can factor all contracted derivatives and express them just as:
\begin{equation}
J_{\mm(s)}^{pl}\approx\sum_l \tilde{\alpha}_{s}^l\left(\frac{\Box}{2}\right)^l J^{\text{can.}}_{\mm(s)}\,,
\end{equation}
assuming the scalar to be massless too and where with $\approx$ we mean on-shell with respect to the scalar.\footnote{At the equation of motion level, the equations of motion for the scalar are just part of the equations and can be used without any problem. The Lagrangian counterpart of this is a redefinition of the scalar field.}
At this point, one can proceed in more than one way. An option is to integrate box by parts under the integral sign so as to get something that is vanishing on-shell unless $l=0$. Another option is to realise that the fact that $\Box$ could be factorised for each individual term in the sum allows to find a redefinition removing the higher-derivative tail: $\delta S=\int\phi\Box\delta\phi$. At the equations of motion level:
\begin{equation}
\Box\phi_{\mm(s)}+\ldots=\left(\frac{\Box}{2}\right)^l J^{\text{can.}}_{\mm(s)}\,,
\end{equation}
this is even more clear since the source terms above precisely have the structure $\Box\delta\phi_{\mm(s)}$ typical of redefinitions and can be removed with:
\begin{equation}
\phi_{\mm(s)}\rightarrow \phi_{\mm(s)}+\frac12\left(\frac{\Box}{2}\right)^{l-1} J^{\text{can.}}_{\mm(s)}\,.
\end{equation}
The latter redefinition would generate cubic and higher order terms in the scalar field contributing to higher contact terms.

In flat space, yet another way to understand the idea described above is to consider the same vertex in momentum space. The higher derivative couplings just discussed would then look simply:
\begin{equation}
\delta^{(d)}(p_1+p_2+p_3)\,(p_2\cdot p_3)^{l}(p_2-p_3)_{\mm(s)}\,\widetilde{\phi}^{\mm(s)}(p_1)\widetilde{\Phi}^*(p_2)\widetilde{\Phi}(p_3)\,.
\end{equation}
The logic is then to realise that up to redefinitions (or equivalently on-shell) the following identity holds:
\begin{equation}\label{momenta}
(p_2\cdot p_3)^{l}=\left[\frac12(p_1^2-p_2^2-p_3^2)\right]^l\approx\left[\frac12(m_1^2-m_2^2-m_3^2)\right]^l\,,
\end{equation}
which indeed tells us that the above coupling vanishes in the massless case arriving to the same conclusions drawn above. Notice however that this would not be true already for massive fields in flat space giving a non-vanishing projection of the higher derivative couplings on the primary canonical ones. The latter observation is relevant, in particular, for AdS space (see below) where indeed the AdS mass of the HS fields is non-vanishing also for massless representations.

All the above possibilities are equivalent. Therefore, in flat space and for massless fields the pseudo-local tail does not have physical effect on S-matrix amplitudes at this order. We stress that the trick here at the S-matrix level is to commute the infinite sum and the integral\footnote{This is assumed to be always possible in a pseudo-local theory.} over space-time. Performing first the integral than the sum one would conclude that, since each individual term vanishes identically on-shell (can be removed by a local redefinition), the corresponding infinite sum also vanishes (can be removed by a redefinition), apart from the $l=0$ term.

The integration by parts trick is a quick and effective way to see what is going on. However, factorizing $\Box^l$ in a higher-derivative current just makes manifest the fact that the above terms can be removed by a field redefinition. This step does not actually require any integration by parts, as it is clear working just at the equations of motion level. To summarise, the above discussion shows that in flat space and for massless fields higher-derivative currents have a vanishing projection onto the primary canonical one \eqref{standardcurrent}. For this reason they can be removed by an admissible pseudo-local redefinition without affecting the coefficient of the primary canonical current.

Even if the situation in flat space appears clear, things change rather quickly as anticipated upon turning on a cosmological constant. The obstruction to the above argument is that it is not anymore true that contracted derivatives can be represented just as $\left(\tfrac{\Box}{2}\right)^l J^{\text{can.}}_{\mm(s)}$. This is due to the fact that the latter and the former expressions differ now by a chain of commutators arising from moving covariant derivatives around. Those produce lower derivative terms of the same type as those in \eqref{nonl} which may give a leftover proportional to the primary canonical current itself \eqref{standardcurrent}. Furthermore, one must recall that massless Fronsdal's fields have a non-vanishing AdS mass (see also the remark above after eq.~\eqref{momenta}):
\begin{subequations}
\begin{align}
\Box\phi_{\mm(s)}&-\nabla_\mm\nabla^\nn\phi_{\nn\mm(s-1)}+\frac12\nabla_\mm\nabla_{\mm}\phi^{\nn}{}_{\nn\mm(s-2)}-m_{\phi}^2\phi_{\mm(s)}+2\Lambda g_{\mm\mm}\phi^{\nn}{}_{\nn\mm(s-2)}=0\,,\\ m_{\phi}^2&=-\Lambda[(d+s-3)(s-2)-s]\,.
\end{align}
\end{subequations}
This time, the terms with derivatives contracted together boil down to a single tensor structure, given just by a \emph{primary} canonical current \eqref{standardcurrent}, plus a remainder:
\begin{align}\label{decomp}
\sum_q \alpha_{s,q}^l\,(g^{\nn\nn})^{l}\nabla_{\mm(s-q)\nn(l)}\Phi^* \nabla_{\mm(q)\nn(l)}\Phi+\mathcal{O}(\Lambda g_{\mm\mm}) \approx C^{(s)}_{l+1}(\alpha) J^{\text{can.}}_{\mm(s)}+J^{\text{impr.}}_{\mm(s)}\,,
\end{align}
where the sign $\approx$ indicates that we are dropping terms proportional to the equations of motion of the scalar.
Above, the $C^{(s)}_l(\alpha)$ coefficients depend on the original $\alpha$ coefficients in a way that we will determine later using various formalisms (unfolding, ambient space, etc.). Notice that the above should be considered as an explicit projection of a given current onto its primary canonical component plus what is orthogonal to it: $J^{\text{impr.}}_{\mm(s)}$ --- certain (pseudo-local) current that is a sum of local improvements. Again the equations of motion for the scalar have been used and are implied by the $\approx$ sign. The flat space result is recovered by noticing that the above $C^{(s)}_l(\alpha)$ coefficients and the AdS masses are, for dimensional reasons, proportional to $\Lambda$.

Making more transparent our use of a nomenclature borrowed from the CFT framework, and quite appropriate here, equation \eqref{decomp} is nothing but the statement that in AdS, higher-derivative terms of the type above, successors, do have a non-vanishing projection onto the \emph{primary} canonical current:
\begin{equation}\label{orto}
\left\langle \sum_q \alpha_{s,q}^l\,(g^{\nn\nn})^{l}\nabla_{\mm(s-q)\nn(l)}\Phi^* \nabla_{\mm(q)\nn(l)}\Phi+\mathcal{O}(\Lambda g_{\mm\mm})\,\right |\left.\vphantom{\sum_q}\,J^{(s)}_{\text{can.}}\right\rangle=C_{l+1}^{(s)}(\alpha)\,.
\end{equation}
The higher derivative currents on the left that we referred to above as \emph{successors} of the primary canonical current should be distinguished from improvements, which would be more properly associated with \emph{descendants} of the primary canonical current with a vanishing projection on it (see the explicit decomposition of eq.~\eqref{decomp}). To summarise, all of this gives a definition of an orthogonal basis for currents, allowing us to define an admissible redefinition at this order, to be \emph{any} pseudo-local redefinition that does not affect the \emph{primary} canonical current sector.

Finally, the fact that everything boils down to a finite number of tensor structures is consistent with the classification of cubic couplings and with the corresponding classification of three-point functions from the CFT side. The link between the above locality definition and the CFT definition is here quite manifest since, as we will show in Section~\ref{Sec: Witten Diagrams}, improvements do not contribute at this order to the computation of CFT correlators.

\paragraph{Locality:}

We are now ready to formulate our locality criterion. Starting from any pseudo-local expression we can reduce cubic couplings to a finite number of tensor structures (primary currents) plus improvements (descendants). Hence, a pseudo-local coupling \eqref{decomp} will boil down to a linear combination of such primary tensor structures multiplied by infinite sums of the type:\footnote{A precursor of this functional class prescription was also given in \cite{Joung:2011ww,Joung:2012fv} in order to present a classification of couplings in the metric-like language.}
\begin{equation}
\left(\sum_{l=1}^{\infty}C^{(s)}_l(\alpha)\right)\int  \, \phi_{\mm(s)} J_{\text{can.}}^{\mm(s)}+\text{Improvements}\,.
\end{equation}
The locality criterion proposed here amounts to the condition that the above series converges:
\begin{equation}\label{loccrit}
\sum_{l=1}^{\infty}C^{(s)}_l(\alpha)<\infty\,,
\end{equation}
while improvement terms (descendants) can be removed by an admissible redefinition (see above) and furthermore do not contribute to the computation of CFT correlators at this order (see Section \eqref{Sec: Witten Diagrams}). The above shows how a pseudo-local tail can be resummed into a primary canonical current piece plus descendants.
We stress that no field-redefinition is involved in this step, while of course improvements can be removed if desired, bringing the result into its local form. Therefore, our locality criterion amounts to a \emph{projection} of the initial current in the orthogonal basis described in eq.~\eqref{orto}.

In the following we shall consider various examples of pseudo-local theories including in particular Vasiliev's equations in 3d and 4d.
So far there are two interesting observations that we anticipate here:
\begin{itemize}
\item Vasiliev's equations in the Schwinger-Fock gauge both in 3d and 4d fail to pass this criterion giving rise to divergent series (see also \cite{Giombi:2009wh}). The possible interpretation of this observation will be discussed later.
\item A factorially damped behaviour for the $\alpha$-coefficients \emph{may not} suffice in general to satisfy the above criterion putting some constraints on couplings based on $\star$-products, as in Vasiliev's equations \cite{Vasiliev:1999ba}, or on similar exponential functions.\footnote{It would be interesting to link the above criterion to analogous results in the context of string theory or non-local gravity models (see \emph{e.g.} \cite{Moeller:2002vx,Biswas:2005qr} for a non-exhaustive list of references).} In view of these considerations one would need to reconsider also the question of locality at the quartic order, that we plan to address in the near future. Finally, the statement that $\star$-product interactions lead to non-local vertices in AdS can be used to constrain the functional ambiguity present in Vasiliev's equations \cite{Vasiliev:1999ba}. The latter is given in terms of a function $f_\star(B)$ of the master zero-form $B(y,z|x)$. One can then argue that it should be reduced to a simple phase $e^{i\theta}$ upon dropping terms of the type $B\star\ldots\star B$ that do not seem to have an interpretation in terms of local vertices.\footnote{See also \cite{Vasiliev:2015mka} for another argument based on a Fefferman-Graham-type expansion which allows similar conclusions.} This is also in agreement with the CFT expectations which forbids any such functional ambiguity, shedding light on a possible bulk mechanism leading to such a result.
\end{itemize}

In the rest of the paper, following the general discussion and ideas sketched above, we are going to present the explicit form of the coefficients $C^{(s)}_l(\alpha)$ and their asymptotic behaviours using various formalisms. First of all we will discuss the generic result in arbitrary dimension using ambient space techniques. Later on we will use the unfolded techniques to compute the above coefficients in the context of 3d and 4d Vasiliev's theories with the aim of comparing with the couplings found in \cite{Kessel:2015kna,Boulanger:2015ova}.


\section{The Generic Case from the Ambient Space Formalism}\label{Sec: Ambient}

Before fixing the dimensionality to 3d or 4d, it is useful to discuss the generic case that can be dealt with quite easily using ambient space techniques.
The ambient space formalism provides an efficient set of tools to explicitly compute the above $C_l^{(s)}$ coefficients in the general case. The whole idea of ambient space is to embed $AdS_d$ into a flat $(d+1)$-dimensional space with appropriate signature. The Fronsdal fields $\phi_{\mm(s)}$ are then embedded into ambient tensor avatars of the type:
\begin{equation}
\Phi_{M(s)} (X)\,,
\end{equation}
subjected to the following tangentiality and homogeneity constraints:
\begin{align}
X^N \,\Phi_{NM(s-1)} (X)&=0\,,& (X\cdot\partial-s+2+\mu)\Phi_{M(s)} (X)&=0\,.
\end{align}
The standard Fierz system takes the general form
\begin{align}
\Box\Phi_{M(s)} (X)&=0\,,& \partial^N\,\Phi_{NM(s-1)} (X)&=0\,,& \Phi^N{}_{N M(s-2)} (X)&=0\,.
\end{align}
Above, the degree of homogeneity $\mu$ is related to the mass entering the AdS Klein-Gordon equation upon reduction as
\begin{equation}
m^2=-\Lambda\left[(\mu-s+2)(\mu-s-d+3)-s\right]\,,
\end{equation}
where massless fields correspond to $\mu=0$.
Cubic current interactions can be then represented in ambient space as:
\begin{equation}
\int_{AdS}\Phi_{M(s)}J^{M(s)}\,,
\end{equation}
which induces a source to the Fronsdal tensor of the type:
\begin{equation}
[\Box-m^2]\Phi^{M(s)}+\ldots=J^{M(s)}\,.
\end{equation}
Above we have used an appropriate ambient covariant $AdS$-measure,\footnote{A convenient choice of this type is given by the following measure \cite{Joung:2011ww,Taronna:2012gb}:
\begin{equation}
\int_{AdS}=\int_{\mathbb{R}^{d+1}}\delta(\sqrt{X^2}-L)\,,
\end{equation} with $\Lambda=-\frac1{L^2}$ which we use hereafter. The latter is however not unique and can be rescaled by powers of $\sqrt{X^2}$ with the effect of changing slightly the basis of couplings after performing the radial reduction.
} while we refer for more details of the formalism to the literature.

This formalism gives a simple rule for the integration by parts of $\Box$ under the integral sign that can be summarised on-shell, and in the TT gauge that suffice for our purposes, as:\footnote{For more details see Appendix B of \cite{Joung:2013nma}.}
\begin{equation}\label{improve}
 \Box J^{M(s)}(X)\sim\Lambda (\Delta+s+d-3)(\Delta-s+2) J^{M(s)}(X)\,.
\end{equation}
Here $\Delta$ is the degree of homogeneity of the current in the ambient coordinate $X$ ($X\cdot\partial J=\Delta J$).

The above formula can be easily derived by contracting the conserved current with an on-shell field and integrating the corresponding density by parts with respect to the given AdS measure. By iterations one then arrives to an explicit formula for the $C^{(s)}_l$ coefficients in this formalism:
\begin{multline}\label{improve2}
 \Box^l J^{M(s)}(X)\sim\Lambda (\Delta-2l+s+d-1)(\Delta-2l-s+4) \Box^{l-1}J^{M(s)}(X)\\=\Lambda^l\left[\prod_{k=1}^l  (\Delta-2k+s+d-1)(\Delta-2k-s+4)\right]J^{M(s)}(X)\,.
\end{multline}
The locality criterion then requires:
\begin{equation}
\sum_l4^l\alpha^{(s)}_l\left(\frac{3-\Delta-s-d}{2}\right)_l\left(\frac{2+s-\Delta}{2}\right)_l<\infty\,,
\end{equation}
where we use the ascending Pochammer symbol $(a)_l=a(a+1)\cdots(a+l-1)$.
From  the above, for given spin and $\Delta$, one can read off the required asymptotic behaviour on the coefficients $\alpha^{(s)}_l$. In the simplest case $\mu=0$, that is relevant for the gauge multiplet in Vasiliev's equations, one finds for $0-0-s$ currents $\Delta=-s-4$ obtaining:
\begin{equation}
\sum_l4^l\alpha^{(s)}_l\left(\frac{7-d}{2}\right)_l\left(3+s\right)_l<\infty\,.
\end{equation}
The above coefficients show how fast the $\alpha^{(s)}_l$ need to go to zero to satisfy the locality criterion. Even a factorially damped behaviour typical of exponential functions and $\star$-products may not be sufficient. Notice that in odd dimensions $d\geq7$ the above ambient basis of improvements collapses so that there is no overlap between higher-derivative currents and the primary canonical ones, as in flat space. This is a particular feature of the above basis and it is sufficient to change slightly the ambient measure by including some factor of $\sqrt{X^2}$ together with the $\delta$-function to modify the basis in such a way that any higher-derivative term has some non-vanishing overlap with primary canonical currents.

To summarise, it is a generic feature of $AdS_d$ that higher-derivative currents have a non-vanishing overlap with primary canonical currents requiring some care when resumming an infinite number of higher-derivative currents.

The generic lesson we can draw from the above result is that any HS interaction that can be expressed in terms of $\star$-products up to possible polynomial damping coefficients, may not pass the above criterion, since $\star$-products typically give just a factorial damping $\frac{1}{l!}$.

In the next section we shall discuss in detail the 3d and 4d unfolded examples commenting explicitly on Vasiliev's equations. Indeed each formalism requires a detailed study of the above coefficients because of the different normalisations of the basis that are used to write down the currents (see also \cite{Gelfond:2010pm} for some discussion of current interactions in the context of 4d unfolded formalism).


\section{3d Unfolding and $0-0-s$ Currents}\label{Sec: 3d}
In this section we introduce the ingredients to compute the $C_l$ coefficients directly in the unfolded language \cite{Vasiliev:1988sa} in order to be able to compare explicitly with 3d Vasiliev's equations and to study their degree of non-locality. For a review of the unfolded formalism and for the explicit currents extracted from Vasiliev's equations we refer to \cite{Kessel:2015kna}.

The system we are interested in is a generic (perturbative) current interaction that, in the unfolded language, can be described by the following equations:
\begin{subequations}
\begin{align}
\adD\omega^{(2)}&=\formJ(C^{(1)},C^{(1)})\,,\\
\tadD C^{(1)}&=0\,,\\
\adD \formJ&=0\,.
\end{align}
\end{subequations}
The above equations are formulated in terms of a zero-form $C(y,\phi|x)$ describing a scalar field, and a master one-form $\omega(y,\phi|x)$ encoding the HS gauge potentials in the frame-like formalism \cite{Vasiliev:1992gr}. The superscript $(n)$ refers to the order in perturbation theory and will be omitted whenever possible. 
The system above is written in terms of appropriate nilpotent derivatives:
\begin{subequations}
\begin{align}
\label{covDer}
\adD&=\nabla-\phi h^{\ga\ga}y^{\vphantom{y}}_\ga\pl_\ga\,,\\
\tadD&=\nabla+\frac{i}{2}\phi h^{\ga\ga} (y_\ga y_\ga-\pl_\ga \pl_\ga)\,,
\end{align}
\end{subequations}
where $\nabla$ is the covariant Lorentz derivative:
\begin{equation}
\nabla=d-\omega^{\ga\ga}y^{\vphantom{y}}_\ga\pl_\ga\,.
\end{equation}
Here, $h^{\ga\ga}$ and $\omega^{\ga\ga}$ are the dreibein and spin-connection of $AdS_3$.
The current $\formJ$ describes the backreaction on the one-form sector that is quadratic in the zero-forms. It is supposed to give stress-tensors built out of the scalar field on the r.h.s. of the Fronsdal equations.

Above we have restricted the attention for simplicity to the HS-algebra $hs(1/2)\oplus hs(1/2)$ with $\lambda=\tfrac12$, where we can use the spinorial language of \cite{Vasiliev:1992gr}. In particular we have an explicit realisation for the HS-algebra $hs(1/2)\oplus hs(1/2)$ in terms of spinorial oscillators $y_\ga$:
\begin{align}
\label{eq:osszillators}
[\hat{y}_\ga,\hat{y}_\gb]&=2i\epsilon_{\ga\gb}\,,
\end{align}
supplemented with a Clifford algebra element $\phi$ ($\phi^2=1$). One then works in terms of functions $\omega(y,\phi|x)$ and $C(y,\phi|x)$ of symbols of the above oscillators using the isomorphism
\begin{equation}
\textrm{sp}_2\simeq \textrm{sl}_2\simeq \textrm{so}(1,2)\,,
\end{equation}
realised through standard $\sigma$-matrices, while contracting all spinorial indices with auxiliary spinorial variables:
\begin{align}
\label{eq:spinref}
V^{a(k)}\,\underbrace{\,\sigma_{a}{}^{\ga\ga}\cdots\sigma_a{}^{\ga\ga}}_{k}\qquad \longleftrightarrow\qquad \sum_k V^{\ga(2k)}y_{\ga(2k)}= V(y)\,.
\end{align}
As anticipated, the zero-form $C(y,\phi|x)$ describes in 3d a complex scalar field with conformal mass together with all of its derivatives:
\begin{equation}
C_{\ga(2n)}\sim(\sigma_{\ga\ga}^\mm\nabla_\mm)^n C(x)\,,
\end{equation}
while the current $\formJ$ is a generic conserved bilinear in the zero forms that can be written as:
\begin{align}
\label{eq:fourierDecomp}
\formJ&= \int d\xi\, d\eta\, \formK(\xi,\eta,y)\, \pC(\xi,\phi|x) \pC(\eta,-\phi|x)\,,
\end{align}
in terms of the Fourier transformed zero form
\begin{align}
\label{eq:fourier}
\pC(y,\phi|x)=\int d\xi \, e^{iy\xi} \, \pC(\xi,\phi|x)\,.
\end{align}
For ease of notation we call both $C(y,\phi|x)$ and its $y$-Fourier transform with the same letter.

Now the main observation is that there is a basis that diagonalises the action of the adjoint derivative $D$ and that disentangles independently conserved subsectors of it \cite{Prokushkin:1999xq,Kessel:2015kna} (pure improvements, primary canonical currents and their successors). This basis will be used to identify the canonical current sector of the backreaction and to find explicit improvement terms to further disentangle all successors from the primary canonical piece.

In order to define such basis we will introduce a convenient generating function technique whose main virtue is to map any tensor operation in the above language into standard ODE.

\subsection{The Contour Integral Representation for the Currents}

The above representation \eqref{eq:fourierDecomp} of the generic 3d current can be decomposed in terms of three basic contractions:
\begin{equation}\label{furierexp}
\formK(\xi,\eta,y)\sim\tilde{\mathcal{K}}(\tau,r,s)=\exp\Big[i\tau\xi^\ga\eta_\ga+is y^\ga\xi_\ga+ir y^\ga\eta_\ga\Big]\,.
\end{equation}
Each of the above terms in the exponents encodes one of the 3 possible contractions among indices belonging to the zero-forms:
\begin{align}
(y\xi)^n(y\eta)^m(\xi\eta)^l\sim C_{\ga(n)\nu(l)}C^{\nu(l)}{}_{\ga(m)}\,.
\end{align}
Furthermore, by defining:
\begin{align}
\omega&=\tau^{-1}\,,& \beta&=(s+r)^{-1}=X^{-1}\,,& \gamma&=(s-r)^{-1}=Y^{-1}\,,
\end{align}
one can rewrite the most general tensorial expression in terms of generating functions of their relative coefficients:
\begin{subequations}\label{nicebasis}
\begin{align}
\formJ^{(0)}&=\ \ \oint (\beta\gamma)\,J^{(0)}\,\tilde{\mathcal{K}}\,,\\
\formJ^{(1)}&=\phi\ \oint h^{\ga\ga}\,\,\left[J^{(1)}_1y{}_\ga y{}_\ga-2i(\beta\gamma)J^{(1)}_2 \, y_\ga\partial_\ga -4(\beta\gamma)^2 J^{(1)}_3\,\partial_\ga \partial_\ga\right]\tilde{\mathcal{K}}\,,\label{nicebasis2}\\
\formJ^{(2)}&=\frac{1}{4}\ \oint H^{\ga\ga}\left[J^{(2)}_1y{}_\ga y{}_\ga-2i(\beta\gamma)J^{(2)}_2 \, y_\ga\partial_\ga -4(\beta\gamma)^2 J^{(2)}_3\,\partial_\ga \partial_\ga\right]\tilde{\mathcal{K}}\,,\\
\formJ^{(3)}&=\frac{\phi}{6}\oint H\, (\beta\gamma)\,J^{(3)}\,\tilde{\mathcal{K}}\,.
\end{align}
\end{subequations}
Here the contour integration is over $\tau$, $X$ and $Y$,
\begin{align}
H^{\ga \ga} = h^{\ga}{}_\sigma \wedge h^{\ga \sigma}\,, && H=H_{\ga\ga}\wedge h^{\ga\ga}\,,
\end{align}
while we use a contour integral representation for the generating functions of relative coefficients (see \cite{Kessel:2015kna} for more details). Above, we have also fixed the freedom in Fierz identities by reducing the ansatz to 3 functions only instead of 6.
The functions
\begin{equation}
J_i^{(k)}(\omega,\beta,\gamma)\sim\sum_{l=1}^\infty k_{i}^{(k)}(l,n,m)\,\omega^{l}\,X^{-n}\,Y^{-m}\,,
\end{equation}
are formal series in $\tau^{-1}$, $X^{-1}$, $Y^{-1}$ encoding the relative coefficients of each tensor structure. Notice that we are basically stripping off the factorial damping associated with the exponential. The factorial damping will appear again after the contour integration has been performed.
The $\sim$ sign just implies that the above formal expansion should be considered as an equivalence class modulo arbitrary polynomial functions in $\tau$, $X$ and $Y$ that vanish upon performing the contour integration (see \emph{e.g.} \cite{Prokushkin:1999xq,Kessel:2015kna}). The $\tilde{\mathcal{K}}$ function can be rewritten in terms of $X$ and $Y$ as:
\begin{equation}
\tilde{\mathcal{K}}(\tau,X,Y)=\exp\Big[\frac{i}{2}\,\tau\,\zeta^+\cdot\zeta^-+\frac{i}{2}\,X\, y\cdot\zeta^++\frac{i}{2}\,Y\, y\cdot\zeta^-\Big]\,,
\end{equation}
where we have defined $$\zeta^\pm_\ga=(\xi\pm\eta)_\ga\,.$$
Remarkably, one can observe that in the above basis both $\nabla$ and $D$ preserve the power of $X$ and $Y$.  It is convenient then to restrict the attention to such orthogonal subsectors of the backreaction:
\begin{equation}
k_i^{(n,m)}(\omega)\sim\sum_{l=1}^\infty k_i(l,n,m)\,\omega^l\, X^{-n} Y^{-m}\,.
\end{equation}
The explicit form of the derivative $D$ as well as other differential operators in terms of the above representation can be found in \cite{Prokushkin:1999xq,Kessel:2015kna}. In the following we are interested in the canonical current sector sitting in two components of the above basis. These are in $k_3(\tau^{-1})$ with possibly two choices $(n=1-2s,m=1)$ and $(n=1,m=1-2s)$ respectively for the $X$ and $Y$ dependence. The latter boil down to a single component upon performing a bosonic projection. Any other choice is related to improvement tensor structures \cite{Kessel:2015kna}. It is easy to see that with this choice of the $X$ and $Y$ dependence all other components of the current vanish identically due to the contour integration, while conservation is trivial for any choice of the function $k_3(\omega=\tau^{-1})$. It is precisely this functional freedom that encodes the higher-derivative tail of improvements present in this sector, while the primary canonical current corresponds just to the choice $k_3(\omega)=\omega$ up to an overall constant --- successors are here represented by higher powers of $\omega$: $\omega^{2n+1}\sim\Box^n$.

On this sub-sectors we can easily write down the representation of the $\adD$ operator as well as the operator that computes the source to the Fronsdal tensor solving the torsion constraint.

\paragraph{$J_3$ with $(n=1-2s,m=1)$:} In this component of the backreaction the generic form of the redefinition (exact current in cohomology) can be given in terms of a function $f(\tau)$ as:
\begin{equation}
\delta J^{(1-2s,1)}_3(\tau)=(1+\tau)^{2s}(1-\tau)^2\partial_\tau\left[(1+\tau)^{1-2s}(1-\tau)^{-1}f(\tau)\right]\,.
\end{equation}
The corresponding source to the Fronsdal tensor can be also easily obtained and expressed in terms of $J^{(2)}_3(\tau)$:
\begin{equation}
J^{\text{Fr.}}_3(\tau)=\frac{1}{2(s-1)}\,(1-\tau)^{2s}(1+\tau)^2\partial_\tau\left[(1-\tau)^{1-2s}(1+\tau)^{-1}J_3^{(2)}(\tau)\right]\,.
\end{equation}
We recall that the above should be considered in the space of functions defined as formal series in $\omega=\tau^{-1}$.

\paragraph{$J_3$ with $(m=1-2s,n=1)$:} In this case the generic form of the redefinition is given in terms of a function $f(\tau)$ as:
\begin{equation}
\delta J_3^{(2)}(t)=(1-\tau)^{2s}(1+\tau)^2\partial_\tau\left[(1-\tau)^{1-2s}(1+\tau)^{-1}J_3^{(1)}(\tau)\right]\,.
\end{equation}
The corresponding source to the Fronsdal tensor can be also easily obtained and expressed in terms of $J^{(2)}_3(\tau)$:
\begin{equation}
J^{\text{Fr.}}_3(\tau)=\frac{1}{2(s-1)}\,(1+\tau)^{2s}(1-\tau)^2\partial_\tau\left[(1+\tau)^{1-2s}(1-\tau)^{-1}J_3^{(2)}(\tau)\right]\,.
\end{equation}
Again, the above should be considered in the space of generating functions defined as formal series in $\omega=\tau^{-1}$.

Before moving on with the analysis of the locality criterion it is useful to have a look at the currents which, upon solving the torsion condition, give the \emph{primary} canonical current as source to the Fronsdal tensor. This current is also pseudo-local \cite{Kessel:2015kna} and the corresponding coefficients will already give us an idea of the required convergence properties that are needed.

The solution to this problem can be found with the above ODE techniques and reads for $(n=1-2s,m=1)$:
\begin{equation}
J_3^{(2)}(\omega)\sim\frac{(1-\omega)^{2s-1}(1+\omega)}{\omega^{2s}}\,\log(1-\omega)\,,\label{CanFron}
\end{equation}
while the corresponding coefficients
\begin{equation}
\alpha_l^{(s)}=\frac{(l-1)!}{(l+2s)!}\,,
\end{equation}
go to zero as
\begin{equation}
\alpha_l^{(s)}\sim \frac{1}{l^{2s}}\,.
\end{equation}
The above behaviour should be confronted with the one obtained from 3d Vasiliev's equations in the Schwinger-Fock gauge \cite{Kessel:2015kna}, given by $\alpha_l^{(s)}\sim\tfrac1{l^3}$.

The above result can be also written down using a more standard notation by performing the contour integrations in $\tau=\omega^{-1}$. The two-form $\formJ$ that is pseudo-local but yields the primary canonical currents as sources to the Fronsdal's equation has a rather simple form:
\begin{align}
\formJ=\int_0^1 dt\,\frac{2-t}{t} H^{\ga\ga}\pl_\ga\pl_\ga \exp i\left[ty(\xi-\eta)-(1-t)\xi\eta\right]C(\xi,\phi|x)C(\eta,-\phi|x)\,.
\end{align}

\subsection{Local Improvements and Functional Class}\label{sec:local_impr}
The above formalism is quite efficient in order to define local improvements within the canonical current sector so as to remove the higher derivative tail as described in Section \ref{Sec: General Crit}.
What we want to do is to find the local redefinition that projects some higher power of $\omega=\tau^{-1}$ ($\omega^{2n+1}\sim\Box^n$), i.e. successors, into the primary canonical current given by a single power of $\omega=\tau^{-1}$:
\begin{equation}
\text{local improvement}=(\sqrt{\Box})^{l-1} J_{\text{can.}}-C_{l}\, J_{\text{can.}}\,,
\end{equation}
where we have used\footnote{Indeed, a pair of contracted derivatives is equivalent to $(\xi\eta)^2$ in the spinorial language, c.f. \eqref{furierexp}.} $\xi^\ga\eta_\ga\sim\sqrt{\Box}$ while fixing our convention for the coefficients $C_l$.
This becomes a simple ODE problem whose solution will determine for us the coefficients $C_{l}^{(s)}$.

What we need to do is to require the particular solution to the following ODE to be a polynomial:
\begin{equation}
\omega^{-2s}(1+\omega)^{2s}(1-\omega)^2\partial_\omega\left[\omega^{2s}(1+\omega)^{1-2s}(1-\omega)^{-1}J_3^{(1)}(\tau)\right]\sim C_l^{(s)}\,\omega -\omega^l\,,
\end{equation}
so as to identify a orthogonal basis for improvements with no overlap with the primary canonical current piece.
The general solution to the above differential equation modulo polynomials can be extracted in integral form as:
\begin{equation}
J_3^{(1)}(\omega)\sim\frac{(1-\omega)^{2s-1}(1+\omega)}{\omega^{2s}}\int_0^\omega dx\,\frac{x^{2s}}{(1-x)^2(1+x)^{2s}}\left[C_l^{(s)}\,x -x^l+p(x^{-1})\right]\,.
\end{equation}
One can then determine $C_l^{(s)}$ enforcing the cancellation of the single poles of the integrand in $x=1$ and $x=-1$. Those indeed, due to the factor multiplying the integral, are the only terms producing non-polynomial logarithmic contributions to the solution. Their absence is equivalent to polynomiality (locality) of $J_3^{(1)}(\omega)$. We thus get the following locality condition for the function $J_3^{(1)}(\omega)$:
\begin{align}
\oint_{x=1}\frac{x^{2s}}{(1-x)^2(1+x)^{2s}}\left[C_l^{(s)}\,x -x^l+p(x^{-1})\right]&=0\,,\\
\oint_{x=-1}\frac{x^{2s}}{(1-x)^2(1+x)^{2s}}\left[C_l^{(s)}\,x -x^l+p(x^{-1})\right]&=0\,,
\end{align}
admitting a unique solution up to a choice of the first coefficient in $p(x^{-1})$:
\begin{equation}\label{Ccoeffs}
C_l^{(s)}=(-1)^l\frac{ s\, (2 s+l)!}{(2 s)! (l+1)!} \Big[2 \,(l+s) \, _2F_1(1,l+2 s+1;l+2;-1)-l-1\Big]+4^{-s} (l+s)\,.
\end{equation}
The above realises the locality criterion in the context of 3d unfolding by computing the explicit projection onto the primary canonical current of each higher-derivative term. The remainder will be given solely by improvements.
A redefinition would than allow us to replace any higher power of $\omega$ with just a linear dependence modulo the above spin-dependent coefficients. The latter redefinitions will otherwise never suffice to remove the primary canonical currents. Notice that as soon as each redefinition above does not affect the computation of Witten diagrams (see Section \ref{Sec: Witten Diagrams}) we can also consider an infinite number of them. On the other hand, allowing for generic pseudo-local redefinitions outside the above class, will affect the Witten diagram result and for this reason would allow for removal of primary canonical currents too \cite{Prokushkin:1999xq,Kessel:2015kna}. The above prescription completely specifies redefinitions that do not affect the Witten diagram computation, as we will explicitly check in Section~\ref{Sec: Witten Diagrams}.

Finally, for later convenience it is quite useful to extract the leading $l$ behaviour of the above coefficients as $l\rightarrow\infty$ since this asymptotic behaviour will constrain the growth of the coefficients in a pseudo-local backreaction. This can be easily extracted from \eqref{Ccoeffs} and reads:\footnote{We can also ask whether the current, which upon solving torsion gives rise to the primary canonical current as source to the Fronsdal tensor, \eqref{CanFron}, is equivalent to the primary canonical current in the frame-like formalism as source to $D\omega$. As one can explicitly check, and as should be expected, the corresponding series converges to $1$ for any $s$, showing in fact their equivalence within our functional class.}
\begin{equation}
C_l^{(s)}\sim l^{2s-1}\,.
\end{equation}

\subsection{Locality and 3d Vasiliev's Equations:} We say that a canonical current, i.e. a primary current with a possibly infinite tower of successors, characterised in the above basis by a function $J_3(\omega)$ admitting an expansion at $\omega=0$ of the type:
\begin{equation}
J_3(\omega)=\sum_{l=1}^\infty \alpha^{(s)}_l \omega^l\,,
\end{equation}
satisfies the locality criterion if the corresponding series that is obtained by rewriting the same coupling by singling out the primary canonical current converges. Namely:
\begin{equation}
\sum_{l=1}^\infty \alpha^{(s)}_l C_l^{(s)}<\infty\,.
\end{equation}
Combining the above criterion with the asymptotic behaviour of the coefficients $C_l^{(s)}$, we are readily able to deduce the corresponding asymptotic behaviour for the pseudo-local backreaction to pass this test:
\begin{equation}
\alpha_l^{(s)}\prec \frac{1}{l^{2s}}\,.
\end{equation}
If the above requirement is not met one would face divergences that would require some analytic continuation in order to link the corresponding backreaction to its primary canonical form. In the case such analytic continuation ($\zeta$-function regularisation for instance) is available we would refer to the corresponding backreaction as non-local. Notice that there might still be a possibility that no such analytic continuation would be available due to cut or pole singularities in the corresponding analytic continuation.

We shall discuss the implications of the above discussion from the AdS/CFT perspective later. Let us conclude this section by applying the above locality criterion to 3d Vasiliev's equations. As described in \cite{Kessel:2015kna} the backreaction of Vasiliev's theory in the Schwinger-Fock gauge has the following asymptotics:
\begin{equation}
\alpha_l^{(s)}\sim \frac1{l^3}\,,
\end{equation}
independently of the spin of the current.
This means that, apart from the $s=1$ case, for which the corresponding sum gives $-\frac{i}{8}$ in the notation of \cite{Kessel:2015kna}, and the $s=2$ case that gives a convergent but not absolutely convergent series resumming to $-\frac{i}{24}$, all other couplings give divergent results. We do not give the explicit form of the divergent series for generic spin because it is not necessary for the discussion and a bit cumbersome. However, we tried standard $\zeta$-function regularisation procedure without succeeding in getting results compatible with HS symmetry \cite{Kessel:2015kna} when using the same regularisation for all spins.

\section{4d Unfolding and Current Couplings}\label{Sec: 4d}
In 4d we can consider the analogue of the problem studied in the previous section (for a review of 4d unfolding we refer to \cite{Vasiliev:1990en,Vasiliev:1999ba,Bekaert:2005vh,Didenko:2014dwa}).
In 4d one can use the isomorphism $so(3,2)\sim sp(4)$ in order to translate space time indices into spinorial $sp(4)$ indices. One then contracts all free indices with symbols of spinorial oscillators $Y_A\sim(y_\ga,\bry_\gad)$ and introduces a master 1-form $\omega(y,\bry|x)$ as a generating function of gauge fields together with a master 0-form $C(y,\bry|x)$ as a generating function of HS Weyl tensors and their derivatives.
Putting together all these ingredients, we are again interested in a generic perturbative unfolded system of equations of the type
\begin{align}
\adD\omega^{(1)}&=H^{\ga\ga}\pl_\ga\pl_\ga C^{(1)}(y,0)+h.c.\,,\\
\adD\omega^{(2)}&=\formJ(C^{(1)},C^{(1)})+H^{\ga\ga}\pl_\ga\pl_\ga C^{(2)}(y,0)+h.c.+\ldots\,,\\
\tadD C^{(1)}&=0\,,\\
\tadD C^{(2)}&=\ldots\,,\\
\adD \formJ&=0\,,
\end{align}
where the $\ldots$ represent further terms which will not be important for the succeeding discussion (see however \cite{Boulanger:2015ova}). The above system reduces to free theory upon setting $\formJ$ to zero. Above the label $(n)$ denotes the perturbative order of the various fields that will be left unspecified unless needed.
The above system is formulated in terms of nilpotent derivatives:
\begin{subequations}
\begin{align}
\tadD &=\nabla+ih^{\ga\gad}(y_\ga \bry_\gad-\pl_\ga \bar{\pl}_\gad)\,,\\
\adD  &=\nabla -h^{\ga\gad}(y_\ga\bar{\pl}_\gad+\bry_\gad{\pl}_\ga)\,,\label{ad}
\end{align}
\end{subequations}
relevant for the description of the corresponding $sp(4)$ modules, while the Lorentz covariant derivative is given by
\begin{align}
\nabla  &=d -\omega^{\ga\ga}y_\ga\pl_\ga-\varpi^{\gad\gad} \bry_\gad\bar{\pl}_\gad\,.
\end{align}
Here above, $h^{\ga\gad}$, $\varpi^{\ga\ga}$ and $\varpi^{\gad\gad}$, are the vierbein and (anti-)self-dual components of the spin-connection of $AdS_4$.
The above system describes the subsector of current interactions that also arise from Vasiliev's equations as studied in \cite{Boulanger:2015ova} (see also \cite{Gelfond:2010pm}). Along lines similar to the 3d case, the generic form of the current can be specified in terms of a kernel:
\begin{equation}
\formJ(C,C)= \int  d^4\xi \, d^4\eta\, \left(H^{\ga\ga} J_{\ga\ga}(Y,\xi,\eta)+
H^{\gad\gad} J_{\gad\gad}(Y,\xi,\eta)\right)\,C(\xi|x) C(\eta|x)\,,
\end{equation}
where we have defined:
\begin{align}
H^{\gad\gad}&=h\fdu{\nu}{\gad}\wedge h^{\nu\gad}\,, & 
H^{\ga\ga}&=h\fud{\ga}{\gnd}\wedge h^{\ga\gnd}\,,
& h\fud{\ga}{\gnd}\wedge H^{\gbd\gnd}&=\hat{h}^{\ga\gbd}\,.
\end{align}
Furthermore, upon considering the following splitting of the adjoint derivative $\adD$ of eq.~\eqref{ad}:
\begin{align}
(\nabla+Q)\omega&=\formJ\,, & Q&=y^\ga h\fdu{\ga}{\gad} \bar{\pl}_\gad +\bry^\gad h\fud{\ga}{\gad}\pl_\ga=Q_++Q_-\,,
\end{align}
one can rewrite the above equation in terms of the components $\omega_{\pm k}$ denoting $\omega^{\ga(s-1\pm k),\gad(s-1\mp k)}$, so that the $k=0$ piece is the HS vielbein. In this way the first two equations read (the one for $\omega_{\pm1}$ splits in two)
\begin{align}
R_0=\nabla \omega_0+Q_+ \omega_{-1} +Q_- \omega_{+1}&=\formJ_0\,,\\
R_{+1}=\nabla \omega_{+1}+Q_+ \omega_{0} +Q_- \omega_{+2}&=\formJ_{+1}\,,\\
R_{-1}=\nabla \omega_{-1}+Q_+ \omega_{-2} +Q_- \omega_{0}&=\formJ_{-1}\,,
\end{align}
and allow to solve for torsion ($R_0$ equations) and get the source to the Fronsdal tensor as we will describe below (see \cite{Boulanger:2015ova} for more details).

\subsection{The 4d Contour Integral Representation for Canonical Currents}

In terms of $y$ and $\bry$ oscillators and using an analogous contour integral representation as the one used in 3d, we have a basis of canonical currents of the type:\footnote{See \emph{e.g.} \cite{Gelfond:2006be} for a conventional generating function thereof.}
\begin{multline}\label{can4d}
J_2^{\text{can.}}=-\oint_{\tau_i,s,r} \left[\beta^2\, g^{(2)}_1(\alpha_1,\alpha_2,\beta,\gamma)H^{\ga\ga}\pl_\ga\pl_\ga+\gamma^2\, g^{(2)}_2(\alpha_1,\alpha_2,\beta,\gamma)H^{\gad\gad}\bar{\pl}_\gad\bar{\pl}_\gad\right]\\\times e^{i(s y\zeta^-+\tau_1\xi\eta+r \bry\brzeta^++\tau_2\brxi\breta)}\,,
\end{multline}
where:
\begin{align}
\alpha_1&=\tau_1^{-1}\,,& \alpha_2&=\tau_2^{-1}\,,&\beta&=s^{-1}\,,&\gamma&=r^{-1}\,,\\
&&\zeta^{\pm}&=\xi\pm\eta\,,&\brzeta^{\pm}&=\brxi\pm\breta\,,
\end{align}
parametrize the four contractions of indices relevant to the canonical current sector in 4d among the six possible ones:
\begin{align}
(y\xi)^n(y\eta)^m(\bry\brxi)^\brn(\bry\breta)^\brm(\xi\eta)^l(\brxi\breta)^{\bar{l}}\sim C_{\ga(n)\nu(l);\gad(\brn)\dot\nu(\bar{l})}C_{\ga(m)}{}^{\nu(l)}{}_{\gad(m)}{}^{\dot{\nu}(\bar{l})}\,.
\end{align}
We should also restrict the attention to the subsector of the backreaction relevant for the canonical current sector at one- and three-form level:
\begin{align}
J_1^{\text{can.}}&=-\oint_{\tau_i,s,r} (\beta\gamma)\, g^{(1)}(\alpha_1,\alpha_2,\beta,\gamma)h^{\ga\gad}\pl_\ga\bar{\pl}_\gad e^{i(s y\zeta^-+\tau_1\xi\eta+r \bry\brzeta^++\tau_2\brxi\breta)}\,,\\
J_3^{\text{can.}}&=-\oint_{\tau_i,s,r} (\beta\gamma)\, g^{(3)}(\alpha_1,\alpha_2,\beta,\gamma)\hat{h}^{\ga\gad}\pl_\ga\bar{\pl}_\gad e^{i(s y\zeta^-+\tau_1\xi\eta+r \bry\brzeta^++\tau_2\brxi\breta)}\,,
\end{align}
respectively. We can then obtain the action of $\adD$ on the above ansatz in closed form:
\begin{align}
\adD g^{(1)}=\begin{pmatrix}
\frac1{4 \alpha_1 \beta }\,\left[\gamma  (\alpha_1 \gamma +\beta ) \partial_\gamma +\alpha_2 \beta  (\alpha_1 \alpha_2+1) \partial_{\alpha_2}\right]\\
\frac{1}{4 \alpha_2 \gamma }\left[\beta  (\alpha_2 \beta -\gamma ) \pl_\beta-\alpha_1 \gamma  (\alpha_1 \alpha_2+1) \pl_{\alpha_1}\right]
\end{pmatrix}g^{(1)}\,,
\end{align}
and
\begin{multline}
D\vec{g}^{(2)}=-\frac{1}{2 \alpha_1 \alpha_2}\Big[\alpha_1 (\alpha_1 \alpha_2+1) (\alpha_1\partial_{\alpha_1})g^{(2)}_1+\alpha_1 \gamma^{-1}  (\gamma -\alpha_2 \beta ) (\beta\partial_{\beta}) g^{(2)}_1+\alpha_1 g^{(2)}_1\\+\alpha_2 (\alpha_1 \alpha_2+1) (\alpha_2\partial_{\alpha_2})g^{(2)}_2+\alpha_2\beta^{-1}  (\alpha_1 \gamma +\beta )(\gamma \partial_{\gamma}) g^{(2)}_2+\alpha_2 g^{(2)}_2\Big]\,.
\end{multline}
that one can explicitly check to be nilpotent.

For the purpose of solving torsion and checking conservation it is also needed to work out the analogous representations for $\nabla$ on the same canonical current space.
The corresponding expressions are given by:
\begin{align}
\nabla g^{(1)}=\frac14\begin{pmatrix}
+\frac{1}{\alpha_1}\left[\alpha_2 (\alpha_1 \alpha_2+1) \partial_{\alpha_2}+\gamma  \partial_{\gamma}\right]\\
-\frac{1}{\alpha_2}\left[\alpha_1 (\alpha_1 \alpha_2+1) \partial_{\alpha_1}+\beta  \partial_{\beta}\right]
\end{pmatrix}g^{(1)}\,,
\end{align}
and
\begin{multline}
\nabla\vec{g}^{(2)}=-\frac{1}{2 \alpha_1 \alpha_2}\Big[\alpha_1^2 (\alpha_1 \alpha_2+1) \partial_{\alpha_1}g_1+\alpha_1 \beta  \partial_\beta g_1+\alpha_1 g_1+\alpha_2 \left(\gamma  \partial_\gamma g_2+\alpha_2 (\alpha_1 \alpha_2+1) \partial_{\alpha_2}g_2+g_2\right)\Big]\,.\nonumber
\end{multline}
In a similar fashion one can also write down the differential operator that computes the source to the Fronsdal tensor upon solving the torsion condition.
For later convenience we give below the operator that computes the source to the Fronsdal tensor given a canonical current in eq.~\eqref{can4d}:
\begin{align}
J_{+1}^{\text{Fr.}}&=+\frac{(\beta\gamma)^{s+1}}{s-1}\,\alpha_2^{-1}\left[\alpha_1 (\alpha_1 \alpha_2+1) \partial_{\alpha_1}k_1(\alpha_1,\alpha_2)+(s+1) k_1(\alpha_1,\alpha_2)+\alpha_2 (s-1) k_4(\alpha_1,\alpha_2)\right]\,,\\
J_{-1}^{\text{Fr.}}&=-\frac{(\beta\gamma)^{s+1}}{s-1}\,\alpha_1^{-1}\left[\alpha_2 (\alpha_1 \alpha_2+1)\partial_{\alpha_2} k_2(\alpha_1,\alpha_2)+(s+1) k_2(\alpha_1,\alpha_2)-\alpha_1 (s-1) k_3(\alpha_1,\alpha_2)\right]\,.
\end{align}
Here we use the notation:
\begin{equation}
\vec{J}_{0}=(\beta\gamma)^{s}\begin{pmatrix}
k_1(\alpha_1,\alpha_2)\\
k_2(\alpha_1,\alpha_2)
\end{pmatrix}\,,
\end{equation}
and
\begin{align}
\vec{J}_{\pm1}=(\beta\gamma)^{s}\left(\frac{\beta}{\gamma}\right)^{\pm1}\begin{pmatrix}
k_3(\alpha_1,\alpha_2)\\
k_4(\alpha_1,\alpha_2)
\end{pmatrix}\,,
\end{align}
for the corresponding components of the current according to the grading induced by $\tfrac12(y\pl-\bry\bar{\pl})$.

As in 3d, before moving to the analysis of locality, it is instructive to find the simplest current within the unfolded formalism which, upon solving torsion, produces the primary canonical current with no higher derivative dressing as source to the Fronsdal tensor. Considering Bel-Robinson type currents of spin $s$ for ease of notation, this problem can be solved with the formalism described above. Starting from a canonical current ansatz written as an exact form:
\begin{subequations}
\begin{align}
\vec{J}_0&=\frac14 \, (\beta\gamma)^s\begin{pmatrix}
+\alpha_2 \,\left[(1+s+\tau ) k(\tau)+\tau  (1+\tau ) k'(\tau)\right]\\
-\alpha_1 \,\left[(1+s+\tau ) k(\tau)+\tau  (1+\tau ) k'(\tau)\right]
\end{pmatrix}\,,\label{J0above}\\
\vec{J}_{+1}&=\, \frac{s}4\,(\beta\gamma)^s\left(\frac{\beta}{\gamma}\right)\begin{pmatrix}
0\\
\tau\,k(\tau)
\end{pmatrix}\,,\\\tau&=\alpha_1\,\alpha_2\,.
\end{align}
\end{subequations}
with arbitrary higher-derivative dressing described by a function $k(\tau)$, the corresponding Fronsdal current can be expressed in terms of a differential operator acting on $k(\tau)$ and reads:
\begin{equation}
\frac{(\beta\gamma)^{s+1}}{4 (s-1)}\Big[\tau  (\tau +1) \left(\tau  (\tau +1) k''(\tau )+k'(\tau ) (2 s+3 \tau +3)\right)+k(\tau ) \left(s^2 (\tau +1)+2 s+(\tau +1)^2\right)\Big]\,.
\end{equation}
Requiring it to be just proportional to a primary canonical current $\tau\,(\beta\gamma)^{s+1}$ fixes the function $k$ to be:
\begin{equation}
k^{(s)}(\tau)=\sum_{l=1}^\infty\,(2l+s+2) \left(\frac{l!}{(l+s+1)!}\right)^2(-\tau)^l\,,
\end{equation}
Hence, one gets an asymptotic behaviour for $k(\tau)$ given by $l^{-2s-1}$ that translates into an asymptotic behaviour for $J_0$ in eq.~\eqref{J0above} given again by $l^{-2s-1}$. As in 3d this behaviour is spin-dependent and converges faster than the Vasiliev equations result $\tfrac1{l^3}$ in the Schwinger-Fock gauge (see \cite{Boulanger:2015ova}). Incidentally the above discussion also gives the exact representative for the primary canonical current. Therefore, we have generalised to 4d the analogous result of \cite{Prokushkin:1999xq,Kessel:2015kna} upon which allowing arbitrary pseudo-local redefinitions it is possible to remove also the primary canonical currents. Below we are going to discuss the functional class of admissible field redefinitions in 4d that avoids this problem.

The above result can be nicely translated back and expressed via homotopy integrals performing the contour integration in \eqref{can4d}. Combining different spins together, we get a simple expression for $\formJ$ that, while still pseudo-local, yields primary canonical currents as Fronsdal currents with some finite coefficients upon solving the torsion constraint. This reads:
\begin{equation}
\formJ=\adD\int_0^1 dt\,dq\,\frac{2-t}{t} h^{\ga\gad}\pl_\ga\bar{\pl}_\gad \exp i\left[ty(\xi-\eta)-(1-t)\xi\eta+q\bry(\brxi+\breta)+(1-q)\brxi\breta\right]\,.
\end{equation}
Here the pole in $t$ is canceled by $\pl_\ga$. As announced, the Fronsdal current contains just the primary canonical current with coefficient:
\begin{equation}
a_{s}=-\frac{1}{2 (s-1)}\,.
\end{equation}
Upon introducing an additional parameter that counts the spin of the current one can change the coefficient above into any other. Such $\formJ$ solves the same problem as studied in \cite{Gelfond:2010pm} where the current interactions involving only primary canonical currents were constructed.

Another example of the backreaction that is pseudo-local but can be localised back to the primary canonical current is:
\begin{align}
\formJ=\adD\int_0^1 dt\,dq\,h^{\ga\gad}\pl_\ga\bar{\pl}_\gad \exp i\left[ty(\xi-\eta)-(1-t)\xi\eta+q\bry(\brxi+\breta)+(1-q)\brxi\breta\right]\,,
\end{align}
which gives the following coefficients for the Fronsdal current:
\begin{align}
\alpha_{s,l}&=\frac{ s }{s-1}\,\left(\frac{l!\, s!}{(l+s+1)!}\right)^2\,.
\end{align}
We see that the above Fronsdal current is pseudo-local. It is important that the dependence on the homotopy parameters yields the beta function that improves the large-$l$ behavior. Thanks to the fast large-$l$ fall-off, the above Fronsdal current can be resummed (see next section) to
\begin{align}
\sum_{l=0}^{\infty}\alpha_{s,l} \, C^s_l=\frac{1}{2(s-1)}\,.
\end{align}
However, such dependence on the spin is still not compatible with the HS symmetry (see \cite{Bekaert:2015tva,Charlotte}). The right spin dependence requires some more work and can be obtained for instance with a clever use of the integral formula for the reciprocal gamma function to define the homotopy integrations.

\subsection{Local Improvements and Functional Class}
As in the 3d case one can translate the projection of higher derivative currents onto primary canonical ones in terms of an ODE. What needs to be done is to identify all local improvements of the schematic form:\footnote{Notice that our convention for $C_l$ is shifted by one unit with respect to the one used in \cite{Boulanger:2015ova}. We choose this convention to make contact with the notation used in the 3d section. To match the convention of \cite{Boulanger:2015ova} it is sufficient to shift $l\rightarrow l+1$ in any formula of this paper.}
\begin{equation}
\text{local improvement}=\Box^{\,l-1} J_{\text{can.}}-C_l \,J_{\text{can.}}\,,
\end{equation}
where we use $(\xi^\ga\eta_\ga)(\brxi^\gad\breta_\gad)\sim\Box$, while projecting out from each higher derivative piece (successor of the primary canonical current) its \emph{primary} canonical component.
One proceeds by studying the generic structure of an improvement term by solving torsion starting with an exact representative for the current. Combining the explicit form of the exact representative with the solution to the torsion equation, the coefficients $C_l^{(s)}$ specifying the aforementioned projection can be then determined by requiring a particular solution to the following differential equation to be polynomial:
\begin{multline}
\frac{1}{4 (s-1)}\Big[\tau  (\tau +1) \left(\tau  (\tau +1) k''(\tau )+k'(\tau ) (n \tau +n+2 s+3 \tau +3)\right)\\+k(\tau ) \left(n (\tau +1) (s+\tau +1)+s^2 (\tau +1)+2 s+(\tau +1)^2\right)\Big]=C_l^{(s,n)}\, \tau +\alpha-\tau ^l\,.
\end{multline}
Above, $n\geq0$ gives the difference between undotted and dotted contracted indices so that when $n=0$ one reduces to the case of Bell-Robinson type currents.

In order to analyze the polynomiality of the particular solution to the above equation we resort to the method of variation of parameters. First of all we need to find the solutions to the homogeneous equation above. These can be conveniently expressed in terms of standard hypergeometric functions as:
\begin{align}
k_1(\tau)&=\frac{\tau ^{n+s} }{(1+\tau )^2}\,{}_2F_1(s,n+s;1+n;-\tau)\,,\label{sol1}\\
k_2(\tau)&=\frac{(-1)^{-n} (-\tau )^{-n} \tau ^{n+s}}{(1+\tau )^2}\,{}_2F_1(s,-n+s;1-n;-\tau )\label{sol2}\,.
\end{align}
The method of variation of parameters gives the following particular solution:
\begin{equation}
k(\tau)=-k_1(\tau)\int d\tau\,\frac{k_2(\tau)\left(C^{(s,n)}_l \tau +\alpha-\tau ^l\right)}{W\,\tau^2(1+\tau)^2}+k_2(\tau)\int d\tau\,\frac{k_1(\tau)\left(C^{(s,n)}_l \tau +\alpha-\tau ^l\right)}{W\,\tau^2(1+\tau)^2}\,,\label{soleq}
\end{equation}
where $W$ is the Wronskian:
\begin{equation}
W=\tau ^{-3-n-2 s} (1+\tau )^{2 s}\,.
\end{equation}
Let us draw the attention on few observations that will allow us to compute the coefficients $C^{(s,n)}_l$:
\begin{itemize}
\item The solution \eqref{sol1} and \eqref{sol2} are independent and both rational functions of $\tau$ with just pole singularities in $\tau=-1$ only for $n\geq s$. They are instead proportional to each other if $n<s$. In the latter case the second independent solution contains logarithmic singularities beyond the pole singularity around $\tau=-1$. Hence, there are two cases to analyze separately.

\item When $n\geq s$ in order to ensure the polynomiality of the above particular solution it is sufficient to require the above two integrands in \eqref{soleq} to have vanishing logarithmic singularity, namely:
\begin{align}
\oint_{\tau=-1}\frac{k_2(\tau)\left(C^{(s,n)}_l \tau +\alpha-\tau ^l\right)}{W\,\tau^2(1+\tau)^2}=&0\,,\\
\oint_{\tau=-1}\frac{k_1(\tau)\left(C^{(s,n)}_l \tau +\alpha-\tau ^l\right)}{W\,\tau^2(1+\tau)^2}=&0\,.
\end{align}
The above equations turn into an algebraic system of equations for $\alpha$ and $C^{(s,n)}_l$ which allows to extract the corresponding value of $C^{(s,n)}_l$:
\begin{equation}
C^{(s,n)}_l=-(-1)^l \frac{(s+l)!}{(s+1)!}\frac{\sum_{k=0}^s\frac{(-1)^k}{\Gamma(1 + l + k - s)}\,F(s,n;k)}{\sum_{k=0}^s\frac{(-1)^k}{\Gamma(2 +k - s)}\,F(s,n;k)}\,,
\end{equation}
where
\begin{equation}\nonumber
F(s,n;k)\equiv \frac{(n-k-1)! }{k! (s-k-1)! (s+n-k-1)! (2s-k)!}\,{}_2F_1(1+k-s,1+k-n-s;1+k-n;1)\,.
\end{equation}
Practically the first equation does not depend on $\alpha$ and allows directly to get the above result.

\item When $n<s$ the solution for $C^{(s,n)}_l$ can be just extracted from
\begin{align}
\oint_{\tau=-1} \frac{k_1(\tau)\left(C^{(s,n)}_l \tau +\alpha-\tau ^l\right)}{W\,\tau^2(1+\tau)^2}=&0\,,
\end{align}
which does not depend on $\alpha$, while $\alpha$ can always be tuned in order to cancel the logarithmic singularity contained into the other solution of the homogeneous equation. Hence, one arrives at:
\begin{equation}
C^{(s,n)}_l=-\frac{(-1)^{l}}{2 (2 s-1) }\frac{n! \,(l+n+s)!}{ s! (s+n-1)!\Gamma(1+l+n-s)}\ {}_3F_2\left(
\begin{matrix}
1-s\quad1+n-s\quad-2 s\\
2-2 s\quad1+l+n-s
\end{matrix}
;1\right)\,.
\end{equation}
\end{itemize}
A few examples for $n=0$, together with the asymptotic behaviour are given below:
\paragraph{$s=2$:}
\begin{equation}
C_l^{(2)}=-\frac{(-1)^l}{2\cdot 3!}\,  l (l+1)^2 (l+2)\,,
\end{equation}
\paragraph{$s=3$:}
\begin{equation}
C_l^{(3)}=-\frac{ (-1)^l}{5!}\, l (l+1) (l+2) (l+3) (l (l+3)+1)\,,
\end{equation}
\paragraph{$s=4$:}
\begin{equation}
C_l^{(4)}=-\frac{(-1)^l}{2 \cdot7!}\, l (l+1) (l+2)^2 (l+3) (l+4) (5 l (l+4)+3)\,,
\end{equation}

\paragraph{$l\rightarrow\infty$ behaviour:}
The generic behaviour can be estimated with analogous techniques observing that in the $l\rightarrow \infty$ limit ($n=0$ for simplicity):
\begin{equation}
_2F_1(s,s;1;-\tau )\sim\sum_{k=0}^{s-1} \frac{\beta_k}{(\tau+1)^{s+k}}\,.
\end{equation}
Setting to zero the pole in $\tau=-1$ in eq. would then require expanding $\tau ^{s} \left(\alpha -\tau ^l+\tau  \chi \right)$ up to order $2s$ around $\tau=-1$. Therefore, the leading behaviour for $l\rightarrow\infty$ is given by:
\begin{equation}
C_l^{(s)}\sim l^{2s}\,.
\end{equation}
The same behaviour holds for $n\neq0$.

\subsection{Locality and 4d Vasiliev's equations}

We should now compare the above asymptotic behaviour with the corresponding behaviour extracted from Vasiliev's equations in the Schwinger-Fock gauge, \cite{Boulanger:2015ova}:
\begin{equation}
\alpha_l^{(s)}\sim\frac1{l^3}\,.
\end{equation}
Surprisingly, the above gives a divergent series for any $s$. This persists also for other couplings beyond the $n=0$ case as soon as the interaction is pseudo-local.
Some examples are the $s=2$ case, for which the overall coefficient of the corresponding primary canonical current reads:
\begin{equation}\label{s2}
-\frac{1}{12} \sum_{n=1}^\infty l\,,
\end{equation}
and the spin-4 and spin-6 cases, where one finds the following primary canonical current overall coefficients:
\begin{align}\label{s4}
s&=4\,,&-\frac{1}{3\cdot 7!}&\sum_{l=1}^\infty\frac{ l (l+1) (l+2)^2 (3 l+11) (5 l (l+4)+3)}{(l+3) (l+4)}\,,\\
s&=6\,,&-\frac{3}{5\cdot 11!}&\sum_{l=1}^\infty\frac{(l+4)!}{(l-1)!}\frac{(l+3) (5 l+28) (7 l (l+6) (3 l (l+6)+19)+20)}{(l+5) (l+6)}\,.
\end{align}
Other spin-$s$ examples can be easily found by combining the $C_l$ coefficients with the coefficients extracted from Vasiliev's theory in \cite{Boulanger:2015ova}.
We then conclude that Vasiliev's equations in the Schwinger-Fock gauge do not give rise to local cubic couplings in the bulk. The main reason for this can be traced back to the fact that couplings are realised in Vasiliev's equations in terms of the $\star$-product with a typical factorial damping $\frac1{l!}$ (that is factorised and always implicit in the contour integral representation above) only improved by an additional $\tfrac1{l^3}$. Notice that even attempting a naive $\zeta$-function regularisation does not seem to give results compatible with HS-symmetry due to the spin-dependent behaviour of the most divergent piece $\sim l^{2s-3}$. Possible interpretation of this result will be discussed in the conclusions. What we can say at this point is that these divergences signal that Vasiliev's equations in the Schwinger-Fock gauge turn out to be a bit more non-local than expected, to the extent that it is not possible to extract from them the coefficient of the primary canonical couplings without specifying a regularisation scheme for the above series \eqref{s2} and \eqref{s4}. Furthermore, let us mention that we have been unable to find any standard $\zeta$-function regularisation compatible with HS-symmetry for all spins. Finally we want to stress here the key difference between our case and the infinite sums appearing in \cite{Giombi:2014yra} is that what needed to be regularised was the sum over spins rather than the sum over contributions from higher-derivative terms.

\section{AdS/CFT and Locality: the 3d Example}\label{Sec: Witten Diagrams}

This section is devoted to correlation function computations in the context of 3d AdS/CFT correspondence (see \emph{e.g.} \cite{Chang:2011vka,Ammon:2011ua,Fitzpatrick:2014vua,Hijano:2015rla,Alkalaev:2015wia} for related results and similar computations). For brevity we concentrate on the 3d case leaving the 4d case for a future publication. The main outcome of this discussion will be to translate the implications of the locality criterion of the previous section in terms of Witten diagrams and corresponding CFT correlators (see also \cite{Heemskerk:2009pn} for some related CFT results). To anticipate the result, the primary canonical current projection mentioned above turns out to match what one would obtain by pulling back in AdS the standard CFT bilinear form, making the CFT nomenclature \emph{primary} current even more appropriate.

\subsection{Scalar Propagators}
Before moving to the detailed amplitude computation it is useful to set all relevant conventions for boundary-to-bulk propagators. The $AdS_3$-space in Poincare coordinates is given by
\begin{align}
h\fud{\ga}{\gb}&=-\frac1{2z}\begin{pmatrix}
dz& dx^+\\
dx^-& -dz
\end{pmatrix}\,, &
\omega\fud{\ga}{\gb}&=-\frac1{2r}\begin{pmatrix}
0& dx^+\\
-dx^-& 0
\end{pmatrix}\,,
\end{align}
with $x^\pm=x\pm t$, which results into
\begin{align}
ds^2=Tr(h \star h)=\frac1{2z^2}(dx^+dx^-+dz^2)\,.
\end{align}
Boundary-to-bulk propagator is a simple Gaussian \cite{Chang:2011mz}:
\begin{align}
C(y)&= K^{\frac12} \exp{\left(\tfrac{i\phi}2 y_\ga F^{\ga\ga}y_\ga\right)}\,,
\end{align}
where $K$ is the Witten propagator
\be
K=\frac{z}{x^+x^-+z^2}\,,
\ee
and the wave vector
\begin{align}
F^{\ga\ga}_0&=\frac{1}{x^+x^-+z^2}\begin{pmatrix}
2z x^+& x^+x^--z^2\\
x^+x^--z^2& -2zx^-
\end{pmatrix}\,,
\end{align}
squares to the unit matrix
\begin{align}
F^{\ga\gb}F\fdu{\gb}{\gc}=\epsilon^{\ga\gc}\,.
\end{align}
Given two points on the boundary, which are specified by some matrices $F_1=(F_1)_\ga{}^\gb$ and $F_2=(F_2)_\ga{}^\gb$, we need to estimate how $C_{\ga(n)\nu(l)}(F_1)\,C\fud{\nu(l)}{\ga(m)}(F_2)$ grows with $l$.
In order to analyse this problem systematically it is convenient to resort again to a generating function representation for the above terms. In the following we introduce the generating function:
\begin{equation}
\mathcal{K}(\tau,s,r)=e^{-i\tau\partial_{u}^\ga\partial_{v}{}_\ga}C(u+sy,\phi)C(v+ r y,-\phi)\Big|_{u=v=0}\,,
\end{equation}
from which one can extract the term with $l$ contractions of indices simply looking at the coefficient of $\tau^l$. Let us notice that every backreaction can be expressed in terms of $\mathcal{K}$.
In the following we are going to compute the above generating function on propagators.
The above computation can be performed explicitly with an integral representation similar to the one used for the Moyal product. To this end, upon performing a Fourier transform on appropriate complex directions we end up with the following integral representation:
\begin{equation}
\mathcal{K}(\tau,s,r)=\frac{1}{\tau^2}\int d^2u\, d^2v\ e^{\tfrac{i}{\tau}u_\ga v^\ga}C(s y+u|\phi)C(r y+v|-\phi)\,.
\end{equation}
It is now straightforward to plug in the Gaussian propagator $C_i(y|\phi)=e^{-\tfrac{i\phi}{2}yF_iy}$ and perform the Gaussian integrals ending up with:
\begin{multline}
\mathcal{K}(\tau,s,r)=[K_1 K_2]^{\tfrac12}\text{det}(1-\tau^2 F_1 F_2)^{-\tfrac12}\\\times\exp\left[-\frac{i\phi}{2}y\left[\frac{1}{1-\tau^2 F_1 F_2}\frac{s}{\tau}(s\tau F_1-\phi r)-\frac{1}{1-\tau^2 F_2 F_1}\frac{r}{\tau}(r\tau F_2-\phi s)\right]y\right]\,.
\end{multline}
For later convenience it is useful to restrict the attention to canonical currents:
\begin{align}
s&=\frac{X}{2}\,,& r&=\frac{X}{2}\,.
\end{align}
Any other choice associated to improvements gives vanishing Witten diagrams regardless the number of derivatives.
We also observe that the above expression has an expansion in terms of $\tau$ of the type:
\begin{equation}
\mathcal{K}(\tau,X)\sim \sum X^{2s} \tau^{l-1} [K_1 K_2]^{\tfrac12+s+\left\lfloor (l-1)/2\right\rfloor}\times\ldots
\end{equation}
where the $\ldots$ are non-singular functions in $x_1^\pm$ and $x_2^\pm$. Assuming that we can compute the CFT correlator summing over each independent contribution for various $l$ we then need the asymptotic expansion of the boundary to bulk propagator $[K_1 K_2]^{\tfrac12+s+\left\lfloor (l-1)/2\right\rfloor}$ as $z\rightarrow 0$. This can be easily extracted by considering the following identity:
\begin{multline}
\int dx^2 \left(\frac{z}{x^+ x^-+z^2}\right)^{\lambda+n} (x^+ x^-)^k=2\pi z^{2-\lambda-n+2k}\int_0^\infty dr \,\frac{r^{2q+1}}{(r^2+1)^{\lambda+n}}\\=\pi z^{2-\lambda-n+2k}\frac{k!\Gamma(n+\lambda-k-1)}{\Gamma(n+\lambda)}\,,
\end{multline}
from which:
\begin{align}\label{expansion}
[K_1 K_2]^{\lambda+s+\hat{l}}&\sim
\pi K_1^{\lambda+s+\hat{l}}\sum_{k=0}^{s+\hat{l}} z^{2-\lambda-s-\hat{l}+2k}\frac{\Gamma(s+\hat{l}+\lambda-k-1)}{k!\Gamma(s+\hat{l}+\lambda)}(\partial_{x^+}\partial_{x^-})^k\delta^{(2)}(x_2)\\
&+\pi K_2^{\lambda+s+\hat{l}}\sum_{k=0}^{s+\hat{l}} z^{2-\lambda-s-\hat{l}+2k}\frac{\Gamma(s+\hat{l}+\lambda-k-1)}{k!\Gamma(s+\hat{l}+\lambda)}(\partial_{x^+}\partial_{x^-})^k\delta^{(2)}(x_1)\,,
\end{align}
where $\hat{l}=\left\lfloor (l-1)/2\right\rfloor$,
while only the leading terms have been kept, discarding trivial contact terms proportional to $\delta^{(2)}(x_1)\delta^{(2)}(x_2)$. As we will see increasing powers of $l$ require to take into account more and more contact terms for the computation of the amplitude. This feature is the counterpart of the convergence behaviour mentioned above and will lead to the same conclusions for locality.

\subsection{$\omega$ Propagator}

We now turn to the discussion of the boundary to bulk propagator for $\omega$. The trick here, since any $\omega$ is pure gauge, is to find the corresponding gauge parameter that would give rise to the correct Brown-Henneaux boundary behaviour \cite{Brown:1986nw,Henneaux:2010xg,Campoleoni:2010zq}.

In our conventions the flat $AdS_3$ connection is embedded into the $\star$-product algebra as follows:
\begin{equation}
W_0=\frac{1}{z}\left[dx^+\, T_{22}+dx^-\, \tilde{T}_{11}+D dz\right]=e^{-\log z D}\left[dx^+\, T_{22}+dx^- \,\tilde{T}_{11}\right]e^{\log z D}+e^{-\log z D} d\, e^{\log z D}\,,
\end{equation}
where we have conveniently defined the generators:
\begin{align}
T_{ii}&=-\frac{i}{4}\left(\frac{1+\phi}2\right)y_i\, y_i\,,& \tilde{T}_{ii}&=-\frac{i}{4}\left(\frac{1-\phi}2\right)y_i \,y_i\,,& D&=-\frac{i}{4}\phi\, y_1 \,y_2\,,
\end{align}
with $D$ the dilatation generator satisfying:
\begin{align}
\left[D,\left(\frac{1+\phi}2\right)y_1\right]_\star&=-\frac12\left(\frac{1+\phi}2\right)y_1\,,\\
\left[D,\left(\frac{1+\phi}2\right)y_2\right]_\star&=+\frac12\left(\frac{1+\phi}2\right)y_2\,,\\
\left[D,\left(\frac{1-\phi}2\right)y_1\right]_\star&=+\frac12\left(\frac{1-\phi}2\right)y_1\,,\\
\left[D,\left(\frac{1-\phi}2\right)y_1\right]_\star&=-\frac12\left(\frac{1-\phi}2\right)y_2\,.
\end{align}
In our conventions:
\begin{align}\label{star}
y_1\star&=y_1+i\partial_{y_2}\,,& y_2\star&=y_2-i\partial_{y_1}\,,\\
\star y_1&=y_1-i\partial_{y_2}\,,& \star y_2&=y_2+i\partial_{y_1}\,.
\end{align}
At the boundary of $AdS_3$ we can then require:
\begin{equation}
D\xi=dx^-\left(-\frac{i}{4}\right)^{s-1}\left(\frac{1+\phi}2\right)y_1^{2s-2}\, 2\pi\mu\,\delta(x_0^+-x_3^+)-dx^+\left(-\frac{i}{4}\right)^{s-1}\left(\frac{1+\phi}2\right)y_2^{2s-2}\frac{2s-1}{(x^+_0-x_3^+)^{2s}}\,,\nonumber
\end{equation}
that upon acting with $e^{-\log z D}$ has the right Brown-Henneaux behaviour for a chiral $\delta$-function source to the spin-s field $\omega$.
Using that the adjoint derivative takes the simple boundary form
\begin{equation}
D\xi=dx^+\left[\partial_++\left(\frac{1+\phi}2\right)y_1\partial_{y_2}\right]\xi+dx^-\left[\partial_--\left(\frac{1-\phi}2\right)y_2\partial_{y_1}\right]\xi\,,
\end{equation}
and requiring the above boundary condition to be matched we get the following simple form for the gauge parameter $\xi$ (see also \cite{Chang:2011vka}):
\begin{equation}
\xi^{(s)}(x_0^\pm;x^\pm_3)=\mu\left(-\frac{i}{4}\right)^{s-1}\left(\frac{1+\phi}2\right)\sum_{n=1}^{2s-1}\frac{1}{(x^+_0-x_3^+)^{n}}y_1^{2s-n-1}y_2^{n-1}\,.
\end{equation}

\subsection{Witten Diagram Computation}
We can now turn to the computation of Witten diagrams using the following trick (see \emph{e.g.} \cite{Chang:2011vka}) to perform the integral over $AdS_3$ that also makes manifest that the 3d theory dynamics is purely from the boundary:
\begin{equation}
\int_{AdS_3}\text{Tr}\left[\omega\star J \right]=\int_{AdS_3}\text{Tr}\left[D\xi^{(s)}\star J\right]=\int_{AdS_3}d \text{Tr}\left[\xi^{(s)}\star J\right]=\lim_{z\to 0}\int_{\partial AdS_3} \text{Tr}\left[\xi^{(s)}\star J\right]_{dz=0}\,.
\end{equation}
Considering now that in Poincar\'e coordinates
\begin{equation}
J\Big|_{dz=0}=H^{\ga\ga}\partial^y_\ga\partial_\ga^y K(\tau)= -\frac{1}{2z^2} dx^-\wedge dx^+ \partial_{y_1}\partial_{y_2}K(\tau)\,,
\end{equation}
and using the $\star$-product formulae \eqref{star} we end up with the following result for the Witten diagram:
\begin{equation}
\lim_{z\to 0}\int dx^- dx^+\frac{\mu}{4z^2}\left(-\frac{i}{4}\right)^{s-1}\sum_{n=1}^{2s-1}\frac{-z^{n-s}}{(x^+_0-x_3^+)^{n}}(i\partial_{y_2})^{2s-n}(-i\partial_{y_1})^{n} K(\tau)\Big|_{y1=y2=0}\,,
\end{equation}
where the dependence on $s$ and $r$ in $\mathcal{K}(\tau)$ has been fixed in order to pick the canonical current sector.
The leftover integral can then be easily evaluated upon substituting the expansion \eqref{expansion} into the above formula and integrating the $\delta$-functions. The main simplification is that all other non-contact terms do not contribute to the limit $z\to 0$.
As a result, considering an expansion in $\tau$ of $\mathcal{K}(\tau)$, we are able to compute all Witten diagrams. The end result is very simple and provides us also with a double check of the computation of local improvements performed in Section \ref{sec:local_impr}. Indeed, calling $W_0$ the result obtained for $\tau=0$, and corresponding to the Witten diagram of a primary canonical current with no higher-derivative contribution, the corresponding $\tau^{l}$ term can be expressed as:
\begin{equation}\label{CFT}
W^{(s)}\Big|_{\tau^{l}}=C^{(s)}_{l+1} W^{(s)}_0\,.
\end{equation}
Here $W_0$ is proportional to the canonical conformal structure up to an overall normalisation:
\begin{equation}
W^{(s)}_0=\pi\,\left(-\frac1{4}\right)^{s}\frac{(s-1)!}{|x_{12}|}\left(\frac{x_{12}^+}{x_{13}^+x_{23}^+}\right)^s\,,
\end{equation}
and the $C^{(s)}_l$ coincide with the coefficients given in eq.~\eqref{Ccoeffs}. This provides a key check also for the bulk definition of bilinear form given in eq.~\eqref{orto} to reproduce the CFT bilinear form \eqref{CFT}. We want to also stress again here that this result is a double-check of the computation of $C^{(s)}_l$ coefficients given in Section \ref{sec:local_impr}. Furthermore, combining the above computation for $W_0^{(s)}$ with the explicit computation of the corresponding coupling constants $g_s=\frac{1}{(2s-2)!}$ at the action level in \cite{Kessel:2015kna} we are also able to match the Witten diagram computation against the CFT one. Taking into account the normalisation included in the generating function $K(\tau)$ one obtains:
\begin{equation}
\left\langle \mathcal{O}(x_1)\bar{\mathcal{O}}(x_2) J_s(x_3)\right\rangle=\pi\,\left(-1\right)^{s}\,\frac{(s-1)!}{|x_{12}|}\left(\frac{x_{12}^+}{x_{13}^+x_{23}^+}\right)^s\,.
\end{equation}
Notice also that in our normalisation the two point function is given by:
\begin{equation}
\left\langle J_s(x_1) J_s(x_2)\right\rangle=\frac{\pi}{4}\,\frac{(2s-1)!}{(x^+_{12})^{2s}}\,,
\end{equation}
which makes our result in complete agreement with the results of \cite{Chang:2011mz} and with the CFT side. Therefore this computation is also a double-check of the coupling constants found in \cite{Kessel:2015kna}, as they precisely match the coupling constants one would extract by requiring the bulk theory to reproduce the CFT correlators.

To summarise, the Witten diagram computation leads to the same conclusions about convergence and hence to the same distinction between non-local and local interactions according to Section \ref{sec:local_impr}. Most importantly the above analysis proves that our criterion for the functional class in HS theory gives a complete prescription for a proper set of non-local redefinitions that \emph{do not} affect the computation of AdS/CFT observables (Witten diagrams in particular). More interestingly, our bulk locality criterion defines a bilinear form on the space of currents \eqref{orto} which agrees with the one on the CFT side \eqref{CFT}. We expect the above results to generalise to higher dimensions too and we defer the explicit 4d Witten-diagram results to a future publication.

To reiterate, the generalised locality condition is tantamount to the requirement that the formal pseudo-local expansion commutes with the integration over AdS space. It might still be however possible that the pseudo-local expansion does not make sense after taking the AdS integral term by term, but it does make sense if one considers the pseudo-local kernel as a whole. In this case the interaction will be considered non-local. This possibility should be understood as the possibility to regularize the sum using some analytic continuation.

\section{Discussions}\label{Sec: Discussions}

The aim of this paper is to fill a gap in the formulation of HS theories by proposing a functional-space for admissible pseudo-local redefinitions. We also give explicit conditions on convergence of the corresponding pseudo-local interactions focusing for convenience on the quadratic order in terms of the equations of motion or cubic order from a Lagrangian perspective. We hope the result of this paper to be useful for a better understanding of the locality issue in the context of HS theories. 

The results presented in this paper provide a criterion to understand the pseudo-local tails arising in HS theories but also present some puzzles in relation with Vasiliev's equations. Indeed the main outcome of this paper is that the Schwinger-Fock gauge usually used to derive unfolded equations from Vasiliev's ones gives rise to \emph{divergent} pseudo-local tails.\footnote{See also \cite{Boulanger:2008tg} for some preliminary discussion regarding the non-local tails arising in Vasiliev's equations.}

Let us stress on the other hand that in \cite{Kessel:2015kna} the cubic action was completely fixed by requiring the associated deformation of gauge symmetries to match the HS algebra structure constants. Unfortunately, similar results can be extended to 4d only for a subset of the vertices for which a pseudo-local tail is found from Vasiliev's equations.

To conclude, we observe that (i) at the level of cubic action the current coupling is unique and any pseudo-local tail can be projected onto its primary canonical component; (ii) recent results \cite{Kessel:2015kna,Boulanger:2015ova,Bekaert:2015tva} as well as general considerations point towards HS theories being ordinary field theories in $AdS$ in the sense that there exists an action with some finite couplings in front of each orthogonal local term, there should be infinitely many of different local couplings though; (iii) HS symmetry favors pseudo-local expressions since it mixes derivatives of the fields. Given (iii) it seems that the most natural vertices consistent with the HS symmetry should still be pseudo-local. This is not in contradiction with (i) and (ii) as one can delocalise any local primary coupling by representing the coefficient $a$ in front of it as a convergent infinite sum $a=\sum_l a_l$. In this way every primary canonical coupling can be replaced with $C^{-1}_l a_l$ times a successor with $l$ pairs of contracted derivatives. Rapid growth of $C_l$ imposes severe restrictions on such delocalised representations. Obviously, any such delocalised coupling can be resummed back into the primary canonical one without producing any infinity.

While a full exhaustive solution to this puzzle is outside the scope of this paper, there are at least two alternatives to consider:
\begin{itemize}
\item Some of the gauge transformations in the Vasiliev theory correspond to non-acceptable field-frames at the level of the physical HS equations of motion. In particular, at this order the gauge ambiguity in Vasiliev's equations amounts to a functional ambiguity that is commonly set to zero, but which plays an important role in determining the physical unfolded equations. Unless the class of such gauge transformations is constrained in some way and at least one point in the domain of admissible gauge transformations is given, one generically ends up with a ``non-local'' HS theory. One would then need to find an appropriate gauge choice tantamount to identifying a \emph{local} HS theory among all possible non-local ones. One option is to find at least one gauge choice in Vasiliev's equations that produces local HS interactions yielding the correct holographic correlation functions compatible with HS symmetry \cite{Colombo:2010fu,Colombo:2012jx,Didenko:2012tv,Didenko:2013bj} without regularising any infinity. At present, however, we can only say that the problem is nontrivial since in the Schwinger-Fock gauge Vasiliev's theory does not obey the locality criterion discussed here, while all naive regularisations of the divergent series we find seem to be incompatible with HS symmetries.

Note that the gauge freedom mentioned above can be also reduced to (not-admissible) field-redefinitions of the physical fields $\omega$ and $C$. Therefore, instead of looking for a physical gauge in Vasiliev's equations we can also look for redefinitions that render the pseudo-local expressions localisable in the sense of eq.~\eqref{loccrit}. However, by looking at $\formJ(C,C)$ we see that there is no way to cure the problem by a simple modification of the coefficients. This is true because of their spin-independent fall-off at $l\rightarrow\infty$. The only resolution would be to redefine the whole pseudo-local current by some exact form $\formJ=\adD\formU$ with a new one $\formJ_{\text{loc.}}$, which can also be represented in $\adD$-exact form, since any $\adD$-closed two-form bilinear in $C$ is exact. Therefore, the redefinition or the gauge choice that solves the problem has to be a quite singular one.

\item Another possible way to attack this problem is to rely on holography by starting from the most general ansatz for the bulk interactions and requiring it to reproduce the correct correlation functions. The outcome of this result will lead us to infer a posteriori the correct gauge-choice. This program however seems to be intractable without using specific features of the theory to be reconstructed, i.e. the HS symmetry for instance.\footnote{See however \cite{Bekaert:2014cea,Bekaert:2015tva} for some recent results along such direction.} In addition one can avoid proliferation of pseudo-local expressions to some extent by choosing the ansatz appropriately and fixing a minimal basis. However, it is important to keep in mind that going to quartic or higher orders in perturbation theory one will face a more complicated version of the same functional class problem analysed in this paper that deserves a close study for a better understanding of HS theories. First of all, it should be analysed what is the minimal basis to study the locality problem at higher orders due to the mixing between current exchange and contact terms. Secondly, a naive extension of the locality condition considered here to the quartic order:
\begin{equation}
\sum_{l,s}\alpha_l^{(s)}J^{(l)}_{\mm(s)}(C,C)\,J^{(l)\,\mm(s)}(C,C)\,,
\end{equation}
with $J^{(l)}_{\mm(s)}(C,C)$ a conserved current with $s+l$ derivatives of the type:
\begin{equation}
J^{(l)}_{\mm(s)}(C,C)\sim \Box^l J^{\text{can}}_{\mm(s)}(C,C)\,,
\end{equation}
suggests the following asymptotic behaviour for the corresponding quartic coefficients in the derivative expansion:
\begin{equation}
\alpha^{(s)}_l\prec\frac{1}{l!^2}\frac{1}{l^{2s+1}}\,.
\end{equation}
which should be verified against explicit computations.
Notice that here we have restored the factorial damping coefficients hidden in our generating function representation of the coefficients.
On top of this, one should also take into account the fact that a redefinition at cubic order will have the effect of mixing current exchange amplitude and contact terms possibly making the contact terms divergent and non-local in some field frame.

For this reason it is quite important to generalise our discussion of locality at higher-orders to be able to quantify along the lines of the discussion carried out here the degree of non-locality of higher-order interactions. We leave a detailed analysis of this problem for the near future.

\end{itemize}

\section*{Acknowledgments}
\label{sec:Aknowledgements}

We would like to thank Nicolas Boulanger, Slava Didenko, Pan Kessel, Gustavo Lucena G\'omez, Rakibur Rahman, Augusto Sagnotti, Charlotte Sleight and Mikhail Vasiliev for useful comments. The research of E.D. Skvortsov and M. Taronna was supported by the Russian Science Foundation grant 14-42-00047 in association with Lebedev Physical Institute.

\begin{appendix}
\section{Notation and Conventions}
\label{app:notation}
\setcounter{equation}{0}

The indices $\mm,\,\nn,\,\ldots=0,\ldots,d$ are covariant AdS indices. Indices $a,\,b,\,\ldots=0,\ldots,d$ are tangent indices at a point on AdS. Indices $M,\,N,\,\ldots=0,\ldots,d+1$ are ambient space indices. Dotted and undotted indices $\ga,\,\gb,\,\ldots=1,2$ and $\gad,\,\gbd,\,\ldots=1,2$ refer to the fundamental and anti-fundamental representations of 4d Lorentz algebra $sl(2,\mathbb{C})$ while in 3d we only have undotted indices of $sl(2,\mathbb{R})$. Finally, the indices $A,\,B,\,\ldots=1,\ldots,4$ label the vectorial representation of $sp(4,\mathbb{R})\sim so(3,2)$.

The symmetrisation convention and index notation goes as follows: symmetrised indices are denoted by the same letter and are assumed to be symmetrised without extra factors. For example, $X_\alpha Y_\alpha$ will stand for $X_{\alpha_1}Y_{\alpha_2} + X_{\alpha_1}Y_{\alpha_2}$, without further normalisation. If symmetric indices sit on the same object we further contract the notation as $T_{\alpha(n)}$. Note that this can lead \emph{e.g.} to expressions of the form $X_\alpha Y_{\alpha(n-1)}$, which should be understood as $X_{\alpha_1}Y_{\alpha_2\dots\alpha_n} + \; (n-1) \; \textrm{terms}$.

All symplectic indices are raised and lowered with the appropriate antisymmetric metric: $\epsilon^{\ga\gb}=-\epsilon^{\gb\ga}$, $\epsilon^{12}=1$ or $C^{AB}=-C^{BA}$ in the $sp(4)$ case that is a block diagonal matrix built from the $sp(2)$ epsilon tensor. By convention, when not displayed explicitly, contraction of indices goes from up-down: $\xi\eta\equiv\xi^\ga\eta_\ga$.

\end{appendix}
\newpage
\begingroup
\setlength{\emergencystretch}{8em}
\printbibliography
\endgroup

\end{document}